\documentclass[aps,pra,floatfix,twocolumn,superscriptaddress]{revtex4-2}
\usepackage{graphicx,hyperref,siunitx,bm}
\usepackage{subfigure}
\usepackage{float}
\usepackage[table]{xcolor}
\usepackage{booktabs}

\hypersetup{
      colorlinks=true,
      citecolor=blue,
      linkcolor=blue,
      urlcolor=blue}
      
\usepackage{amsmath}
\usepackage{amsfonts}
\usepackage{amssymb}
\usepackage{braket}
\usepackage{times}
\usepackage{scalerel}

\usepackage{tikz}
\usetikzlibrary{svg.path}

\usepackage{upgreek}
\usepackage{algpseudocode}
\usepackage{algorithm}

\usepackage{cancel}

\definecolor{ESS}{rgb}{0.392, 0.627, 0.902}   % color{1}
\definecolor{LSS}{rgb}{0.314, 0.549, 0.824}   % color{2}
\definecolor{LEL}{rgb}{0.580, 0.753, 0.910}   % color{3}
\definecolor{TTS}{rgb}{0.949, 0.541, 0.259}   % color{4}
\definecolor{TL}{rgb}{0.988, 0.816, 0.667}    % color{5}
\definecolor{LL}{rgb}{0.941, 0.941, 0.941}    % color{6}
\definecolor{LTS}{rgb}{0.902, 0.902, 0.902}   % color{7}

\graphicspath{{figures/}}
\newcommand\wordcount{\input{|"texcount -inc -sum -0 -template={SUM} \jobname.tex"}}

\newcommand{\tJ}{$t\text{--}J~$}

\newcommand{\myref}[2][]{Fig.\,\hyperref[#2]{\ref*{#2}#1}}
\newcommand{\Myref}[2][]{Figure~\hyperref[#2]{\ref*{#2}#1}}

\definecolor{orcidlogocol}{HTML}{A6CE39}
\tikzset{
  orcidlogo/.pic={
    \fill[orcidlogocol] svg{M256,128c0,70.7-57.3,128-128,128C57.3,256,0,198.7,0,128C0,57.3,57.3,0,128,0C198.7,0,256,57.3,256,128z};
    \fill[white] svg{M86.3,186.2H70.9V79.1h15.4v48.4V186.2z}
                 svg{M108.9,79.1h41.6c39.6,0,57,28.3,57,53.6c0,27.5-21.5,53.6-56.8,53.6h-41.8V79.1z M124.3,172.4h24.5c34.9,0,42.9-26.5,42.9-39.7c0-21.5-13.7-39.7-43.7-39.7h-23.7V172.4z}
                 svg{M88.7,56.8c0,5.5-4.5,10.1-10.1,10.1c-5.6,0-10.1-4.6-10.1-10.1c0-5.6,4.5-10.1,10.1-10.1C84.2,46.7,88.7,51.3,88.7,56.8z};
  }
}

\newcommand\orcidicon[1]{\href{https://orcid.org/#1}{\mbox{\scalerel*{
\begin{tikzpicture}[yscale=-1,transform shape]
\pic{orcidlogo};
\end{tikzpicture}
}{|}}}}

\setcitestyle{super}

\begin{document}

\title{Topology meets superconductivity in a one-dimensional $t$\textnormal{--}$J$ model of magnetic atoms}

\author{Leonardo Bellinato Giacomelli\,\orcidicon{0009-0004-9246-946X}} 
\affiliation{Universität Innsbruck, Fakultät für Mathematik, Informatik und Physik,
Institut für Experimentalphysik, 6020 Innsbruck, Austria}
\author{Thomas Bland\,\orcidicon{0000-0001-9852-0183}}
\affiliation{Universität Innsbruck, Fakultät für Mathematik, Informatik und Physik,
Institut für Experimentalphysik, 6020 Innsbruck, Austria}
\affiliation{Division of Mathematical Physics and NanoLund, Lund University, SE-221 00 Lund, Sweden}
\author{Louis Lafforgue\,\orcidicon{0009-0000-8234-6328}}
\affiliation{Universität Innsbruck, Fakultät für Mathematik, Informatik und Physik,
Institut für Experimentalphysik, 6020 Innsbruck, Austria}
\author{Francesca Ferlaino\,\orcidicon{0000-0002-3020-6291}}
\thanks{Correspondence should be addressed to \href{mailto:francesca.ferlaino@uibk.ac.at}{francesca.ferlaino@uibk.ac.at}}
\affiliation{Universität Innsbruck, Fakultät für Mathematik, Informatik und Physik,
Institut für Experimentalphysik, 6020 Innsbruck, Austria}
\affiliation{Institut für Quantenoptik und Quanteninformation, Österreichische Akademie der
Wissenschaften, 6020, Innsbruck, Austria}
\author{Manfred J. Mark\,\orcidicon{0000-0001-8157-4716}}
\email{manfred.mark@uibk.ac.at}
\affiliation{Universität Innsbruck, Fakultät für Mathematik, Informatik und Physik,
Institut für Experimentalphysik, 6020 Innsbruck, Austria}
\affiliation{Institut für Quantenoptik und Quanteninformation, Österreichische Akademie der
Wissenschaften, 6020, Innsbruck, Austria}
\author{Luca Barbiero\,\orcidicon{0000-0001-9023-5257}}
\email{luca.barbiero@polito.it}
\affiliation{Institute for Condensed Matter Physics and Complex Systems, DISAT, Politecnico di Torino, I-10129, Torino, Italy}

\begin{abstract}
Strongly interacting fermions represent the key constituent of several intriguing phases of matter. However, due to the inherent complexity of these systems, important regimes are still inaccessible. Here, we derive a realistic and flexible setup based on ultracold magnetic lanthanide atoms trapped in a one-dimensional optical lattice. Leveraging their large magnetic moments, we design a fermionic $t$--$J$ model with independently tunable hopping, spin-spin couplings, and onsite interaction. Through combined analytical and numerical analysis, we uncover a variety of many-body quantum phases--including superconducting and topological states. Crucially, in the regime of attractive onsite interaction we reveal that topology and superconductivity coexist, thus giving rise to an exotic state of matter: a topological triplet superconductor. We also outline a practical protocol to prepare and detect all discovered phases using current experimental techniques. Our results establish an alternative and powerful route for a deeper understanding of strongly interacting fermionic quantum matter.
\end{abstract}

\date{\today}
\maketitle
The discovery of some of the most fascinating quantum phenomena such as magnetism \cite{assa,Sachdev2008}, superconductivity \cite{dagotto1994,lee2006}, and topology \cite{Leijnse_2012,Sato2016}, has put strongly interacting fermionic systems at the forefront of research in modern quantum physics. However, the intrinsic complexity of such systems challenges a complete understanding of relevant many-body regimes. In this regard, ultracold fermionic atoms in optical lattices \cite{Lewenstein2012,Gross2017} have emerged as a powerful means of unraveling the properties of a wide variety of complex states of matter, ranging from Mott \cite{Jordens2008}, flavor-selective Mott \cite{Tusi2022}, magnetic \cite{Mazurenko2017,shao2024,lebrat2024}, and topological \cite{Sompet2022,Walter2023} insulators to conducting phases characterized by bad metallic transport \cite{Brown2019}, non-local pairing \cite{Hartke2023}, finite Hall response \cite{Zhou2023}, stripe formation \cite{Bourgund2025}, and pseudogap behavior \cite{chalopin2024}. Notably, the \textit{trait d’union} of these spectacular achievements is the iconic Fermi-Hubbard model \cite{Hubbard1963}, which, in the strongly interacting regimes, is accurately described by the equally celebrated \tJ Hamiltonian \cite{Chao1978}. This paradigmatic model captures the dynamics of fermions subject to a weak and isotropic magnetic coupling induced by large onsite interactions. Crucially, the strongly repulsive regime, in which double occupancy is energetically suppressed, constitutes the only configuration available to alkali atoms for the realization of \tJ Hamiltonians \cite{Hirthe2023,Chalopin2025,Bourgund2025}. Further theoretical \cite{Gorshkov2011} and experimental \cite{carroll2024,douglas2024} developments have enlarged the scope to $t$--$J$ models with large and anisotropic spin-spin interactions. However, they also remain limited to the regime characterized by strong onsite repulsion.

In this paper, we perform a substantial step beyond current available models. In particular, we derive a \tJ model featuring strong and anisotropic spin–spin interactions and where the formation of a significant fraction of double occupancy is energetically permitted. As we show, this configuration becomes accessible in ultracold systems of magnetic lanthanide atoms, where advanced manipulation techniques of the large spin manifold~\cite{Claude2024} can be integrated with the established capability to independently tune onsite interactions~\cite{Chin2010fri,Baier2018roa}.
Importantly, our analytical and numerical analysis unveil that the ground state of this novel Hamiltonian is characterized by some of the most intriguing, yet experimentally unrealized, states of matter: one-dimensional superconductors, a topological liquid, and a topological superconductor, all within experimentally accessible regimes. Finally, we derive rigorous state preparation and probing protocols that pave the way for experimental realization and, more broadly, a deeper understanding of states of matter characterized by topological and/or superconducting order.

\section*{Results}
\subsection*{Model derivation}
Our setup involves $N$ fermionic magnetic atoms of either erbium or dysprosium, trapped in an effectively one-dimensional optical lattice of length $d_x L$, where $d_x$ is the lattice spacing. As widely explored in ultracold atom experiments, the one-dimensional regime can be achieved either by using a three-dimensional lattice with large depths along two directions to suppress tunneling, or by employing a quantum gas microscope combined with established layer-cleaning techniques~\cite{Gross2021}.
In analogy with bosonic realizations \cite{laupretre2025}, this system can be accurately described by the following effective Hamiltonian
\begin{align}\label{efhm}
    H =  &-t\sum_{i,m_F}\left(c_{i,m_F}^{\dagger} c_{i+1,m_F}+ h.c\right) \nonumber \\
    &+\frac{U}{2} \sum_{i,m_F\neq m_F'} n_{i,m_F} n_{i,m_F'}
\end{align}
\begin{align}
    &+\sum_{i<j} V_{i,j} \left[F_i^{z}F_{j}^{z}-\frac{1}{4}(F_i^{+} F_{j}^{-}+F_i^{-} F_{j}^{+})\right] \ ,\nonumber
\end{align}
where $m_F$ denotes the projection along the quantization axis of the total angular momentum $F$, which reaches the notably large values of $F_\text{Er} = 19/2$ for erbium and $F_\text{Dy} = 21/2$ for dysprosium. The fermionic operators $c_{i,m_F}^\dagger$ ($c_{i,m_F}$) create (annihilate) an atom with spin projection $m_F$ at lattice site $i$, while $t$ and $U$ represent the tunneling amplitude and onsite interaction energy, respectively. We underline that higher-spin $t\text{--}J$ models with preserved rotational symmetry can also be realized in SU(N) systems~\cite{Manmana2011SUN}. Unlike in standard Fermi-Hubbard implementations with alkaline atoms~\cite{Esslinger2010}, $U$ also includes a contribution from the magnetic dipole–dipole interaction (DDI), thus offering a further control parameter. This aspect, combined with standard techniques~\cite{Chin2010fri,Baier2018roa} allows for the exploration of various regimes of onsite interaction. The spin operators $F_i^z$ and $F_i^\pm$ capture long-range spin–spin couplings between sites $i$ and $j$, with interaction strength given by $V_{i,j} = V(1 - 3\cos^2\theta)/(|i - j|)^3$, where $V$ is a fixed amplitude determined by the atomic species, and $\theta$ is the angle between the dipole polarization direction and the lattice axis, see Methods.
\begin{figure}[h]
    \centering
    \includegraphics{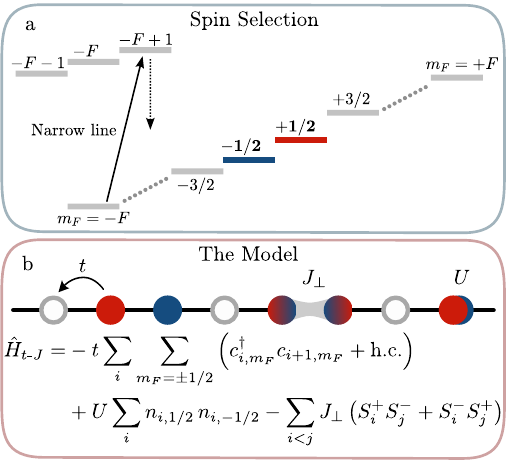}
    \caption{\textbf{Illustration of the model}. (a) The spin selection mechanism. The atoms are pumped from the ground state $m_F=-F$ to the target states via transfer through an long lived clock-like state at  $\lambda_{\rm Er} = 1299$\,nm and $\lambda_{\rm Dy} = 1001$\,nm for Er and Dy respectively  \cite{Claude2024}. Due to hyperfine splitting,  once initialized into $\bra{F, m_F=\pm 1/2}$ the atoms form an isolated spin-1/2 system. (b) The model. 
    The Hamiltonian we derive and analyze in this work comprises three terms: the hopping term $t$, the onsite interaction $U$ and the spin-spin interaction $J_\perp$. In our illustration, the states $m_F=-1/2$ and $m_F=+1/2$ are denoted by the colors blue and red, respectively. The accompanying illustration highlights these processes: the hopping term $t$ is depicted by a black arrow, the spin-spin interaction ($J_\perp$) is represented by the exchange of colors (blue to red and vice versa), and the onsite interaction ($U$) is illustrated as acting on pairs of atoms with different spin state on a single lattice site.}
    \label{fig1:illustration of system}
\end{figure} 

Notably, recent work has shown that the large tensorial polarizability in magnetic lanthanides can be used to perform spin-selective engineering of the quadratic light shift~\cite{Claude2024,Lecomte2025}. This allows to isolate and restrict the system's dynamics only to the $m_F = \pm 1/2$ states, while putting out of resonance all the other states, as illustrated in Fig.\,\ref{fig1:illustration of system}(a).
This reduction allows to rewrite Hamiltonian~\eqref{efhm} as a generalized spin-1/2 $t$--$J$ model
\begin{align}\label{model}
H_{t\text{--}J} =\ &-t \sum_{i} \sum_{m_F=\pm1/2} \left(c_{i,m_F}^{\dagger} c_{i+1,m_F} + \text{h.c.} \right)\nonumber \\
&+ U \sum_i n_{i,1/2} n_{i,-1/2} \\
&+ \sum_{i<j} \left[ J_z S_i^{z} S_j^{z} - J_\perp \left( S_i^{+} S_j^{-} + S_i^{-} S_j^{+} \right) \right],\nonumber
\end{align}
with the effective spin-1/2 operators defined as $S_i^{z} = (n_{i,1/2} - n_{i,-1/2})/2$, $S_i^{+} = c_{i,1/2}^{\dagger} c_{i,-1/2}$, and $S_i^{-} = c_{i,-1/2}^{\dagger} c_{i,1/2}$, acting within the constrained local Hilbert space of the projected total angular momentum, i.e.\,$|0,0\rangle$, $|0,1/2\rangle$, $|-1/2,0\rangle$, and $|-1/2,1/2\rangle$, see Fig.\,\ref{fig1:illustration of system}(b). It is fair to mention that for $U\gg t$, i.e. where the state $|-1/2,1/2\rangle$ is prohibited, Eq. \ref{model} shares similarities with polar molecules setups \cite{Gorshkov2011}. Importantly, in our derivation the reduction of the spin space leads to a significant enhancement of the spin-flip processes $J_\perp = \frac{1}{4} \gamma V_{i,j}$, with $\gamma = F(F+1) + \frac{1}{4}$~\cite{Patscheider2020cde}, in comparison to the Ising magnetic term $J_z = \frac{1}{4} V_{i,j}$. As a consequence, this results in a boost of $J_\perp$ by approximately two orders of magnitude for both erbium ($\gamma_\text{Er} = 100$) and dysprosium ($\gamma_\text{Dy} = 121$), thereby allowing us to neglect the much smaller $J_z$ term. For a realistic lattice spacing $d_x = 266\,$nm~\cite{dePaz2013nqm,Baier2016ebh,Baier2018roa,Lepoutre2019ooe,Patscheider2020cde,Su2023dqs}, it is possible to accurately derive the relevant Hamiltonian parameters--$J_\perp$, $U$, and $t$--for both Er and Dy by integrating over the lowest-band Wannier functions~\cite{Baier2016ebh}, see Methods. Table~\ref{tab:params} presents these values for representative combinations of model parameters, as will become clear in the following sections. 

\begin{table}[b]
\caption{Calculated Hamiltonian parameters for erbium (Er) and dysprosium (Dy) atoms in a one-dimensional optical lattice with spacing $d_x = 266\,$nm. Listed for each configuration are the lattice depth $s_x$ (in recoil energies $E_R$), the angle of the magnetic filed relative to $x$ $\theta$, and scattering length $a_s$ (in Bohr radii $a_0$); followed by the tunneling amplitude $t/h$ (in Hz), the spin-flip interaction strength $J_\perp/t$, the onsite interaction $U/t$; and finally the many-body phase associated with the parameters.}
\label{tab:params}
\begin{tabular}{l @{\hskip 8pt} ccc @{\hskip 12pt} ccc @{\hskip 10pt} c}
\toprule
Atom & $s_x$ (E$_R$) & $\theta$ & $a_s$ ($a_0$) & $t/h$ (Hz) & $J_\perp/t$ & $U/t$ & Phase \\
\midrule
Er & 21 & $90^\circ$ & 0.3  & 8.8  & +1.0 & 1.0  & TL \\
Er & 24 & $90^\circ$ & -0.3 & 5.2  & +1.7 & -1.5 & TTS \\
Er & 19 & $0^\circ$  & -1.5 & 12.6 & -1.4 & -3.0 & LSS \\
Er & 24 & $0^\circ$  & 0.5  & 5.2  & -3.4 & 2.6  & ESS \\
\midrule
Dy & 18 & $90^\circ$ & 0.6  & 15.8 & +1.1 & 1.0  & TL \\
Dy & 20 & $90^\circ$ & -0.6 & 10.9 & +1.6 & -1.5 & TTS \\
Dy & 16 & $0^\circ$  & -3.0 & 23.4 & -1.6 & -3.2 & LSS \\
Dy & 21 & $0^\circ$  & 1.0  & 9.1  & -3.9 & 3.0  & ESS \\
\bottomrule
\end{tabular}
\end{table}

Based on these considerations, two key features of $H_{t\text{--}J}$ are particularly important to highlight. First, unlike alkaline atoms, the strength of the spin–spin interaction can be entirely decoupled from both the tunneling amplitude $t$ and the onsite interaction $U$. Second, the condition $J_\perp \gg J_z$ implies that the effective $SU(2)$ spin-rotational symmetry is explicitly broken, thus phases of matter where such symmetry does not hold can take place without violating the Mermin-Wagner theorem \cite{Mermin1966}. These characteristics stand in stark contrast to the conventional derivation of the $t$–$J$ model, which is obtained in the strong-coupling limit $t \ll |U|$ of the Fermi-Hubbard model~\cite{Chao1978}. In that case, the magnetic interaction $J$ is necessarily isotropic, preserving the $SU(2)$ symmetry, and its magnitude is constrained to $J \ll t$ since it scales as $t^2 / |U|$. 
Notably, this is the precise scenario occurring in experimental setups using alkaline atoms in optical lattices~\cite{Hirthe2023,Bourgund2025,Chalopin2025}.
More experimental flexibility has recently been achieved in dipolar lattice systems~\cite{carroll2024,douglas2024}, where the $SU(2)$ symmetry is also explicitly broken and the magnetic interaction is decoupled from both $t$ and $U$. However, as mentioned, these realizations are restricted to regimes with very large $U/t$, which strongly suppress double occupancy. As a consequence, the local Hilbert space is effectively constrained to \,$|0,0\rangle$, $|0,1/2\rangle$ and $|-1/2,0\rangle$ and regimes with intermediate interaction strengths are inaccessible. It is also worth emphasizing that for magnetic lanthanide atoms, the derivation of $H_{t\text{--}J}$ is completely independent of the average particle density $\bar{n} = N / L$. In contrast, alkaline-atom implementations require that $\bar{n} \neq 1$, while polar molecule experiments typically operate at much lower densities. Taken together, these features underscore the remarkable versatility of our proposed setup. In the following, we show that this tunability is accompanied by a variety of interesting many-body phases.

\subsection*{Topology and superconductivity in the low-energy regime}
We are interested in the ground state properties of Eq.\,\eqref{model} in the regime $\bar{n}<1$ and vanishing total magnetization, $S^z_\text{tot} = \sum_i S_i^z = 0$.\footnote{We expect our results to remain stable in the sector of total magnetization $S^z_\text{tot} =\pm1$. For higher/lower magnetization phase separated regimes, should occur. Here the system presents a region of zero total magnetization where the phase that we discuss persist, and one or more fully polarized Luttinger liquid regions~\cite{Montorsi2020}.}
A first insight can be unveiled by employing the bosonization technique~\cite{Giamarchi2003qpi, Gogolin1998}. This method, which is valid in the low energy regime of $H_{t\text{--}J}$, i.e. when $U, J_\perp \ll t$, allows mapping (see Methods) the microscopic Hamiltonian in Eq.\,\eqref{model} onto a Sine-Gordon model $\mathcal{H}_\text{SG} = \mathcal{H}_c + \mathcal{H}_s$, with decoupled charge and spin sectors, where
\begin{align}\label{Hamiltonian bosonized spin}
    \mathcal{H}_{c,s} = &\frac{v_{c,s}}{2} \int \text{d}x \left[\frac{1}{K_{c,s}} \left(\partial_x \phi_{c,s}(x)\right)^2 + K_{c,s} \left(\partial_x \theta_{c,s}(x)\right)^2 \right. \nonumber \\
    &\left.+ \frac{2g_{c,s}}{d_x^2} \cos\left(\sqrt{8\pi}\, \phi_{c,s}(x)\right)\right] \,.
\end{align}
The phase diagram of Eq.\,\eqref{Hamiltonian bosonized spin} is governed by the behavior of the dual bosonic fields $\phi_{c,s}$ and $\theta_{c,s}$, obtained by replacing the creation/annihilation operators in Eq.\,\eqref{model} with continuous fermionic fields and then expressing these in terms of bosonic fields (see Methods). The labels $c$ and $s$ refer to the charge and spin sectors respectively. Upon a renormalization group (RG) analysis~\cite{Gogolin1998}, the bosonic fields depend on the interaction coupling $g_s = \frac{1}{2\pi^2 v_F}(U + 2 J_\perp)$ and the Luttinger parameters $K_{c,s}$
\begin{align}
    K_{c,s} = 1 \mp \frac{1}{2\pi v_F} \left(U + 2J_\perp \cos(\pi \bar{n})\right)\,,
\end{align}
with corresponding excitation velocities $v_{c,s} = v_F / K_{c,s}$, where the Fermi velocity $v_F = 2 t d_x \sin(\pi \bar{n}/2)$ and $d_x$ is the lattice spacing. Specifically, this RG analysis allows us to determine the relevance of the cosine term in Eq.\,\eqref{Hamiltonian bosonized spin}. This term becomes relevant when the corresponding field $\phi_{c,s}$ is pinned to a constant value, and a finite gap in the $c,s$ excitation spectrum develops (see Methods). Importantly, in the regime $\bar{n} < 1$, the charge sector remains gapless, as the cosine term in $\mathcal{H}_c$ is always irrelevant--i.e.~$\phi_c$ does not pin. Within this analysis, we also know that $\phi_s$ can pin to the specific values ${0, \pm\sqrt{\pi/8}}$~\cite{Nakamura2000,Giamarchi2003qpi,Barbiero2013hho,baldelli2024}. As a consequence, a finite spin gap
\begin{align}
\Delta_s = \lim_{L \to \infty} \left[ E(S_{\text{tot}}^z = 1) - E(S_{\text{tot}}^z = 0) \right]
\end{align}
develops, where $E(S_{\text{tot}}^z)$ denotes the ground-state energy at fixed total magnetization. Physically, $\Delta_s$ corresponds to the energy cost of flipping a single spin. The different pinning values of $\phi_s$ are associated to gapped phases with different properties. In particular, when the field is pinned at $\phi_s = 0$, the system ordering is uniquely captured by the long-range order (LRO) of the parity correlation function 
\begin{equation}\label{parity}
   O_\text{P}^{s} (r)= \langle {\rm e}^{2\pi \imath \sum_{l<r}S_l^z}\rangle,
\end{equation}
which, in its bosonized form \cite{Montorsi2012,Barbiero2013hho}, is written as
\begin{equation}\label{POP}
O_\text{P}^s(r)\sim \langle \cos(\sqrt{2\pi}\phi_s (0))\cos(\sqrt{2\pi}\phi_s (r))\rangle\,.
\end{equation}
Importantly, phases of matter with such a feature are characterized by the formation of local or nonlocal pairing of fermions with opposite spin~\cite{Barbiero2013hho, Montorsi2012}. On the contrary, upon bosonizing the string correlation function 
\begin{align}\label{SOP}
O_\text{S}^{s}(r) = \left\langle S_j^z e^{2\pi \imath \sum_{l=j+1}^{j+r-1} S_l^z} S^z_{j+r} \right\rangle,
\end{align}
it is possible to obtain the following expression~\cite{Montorsi2012,Barbiero2013hho,Fazzini2019iif}
\begin{align}\label{SOPb}
O_\text{S}^s(r) \sim \left\langle \sin\left(\sqrt{2\pi} \phi_s(0)\right) \sin\left(\sqrt{2\pi} \phi_s(r)\right) \right\rangle\,.
\end{align}
Thus, it is straightforward to understand that when $\phi_s = \pm\sqrt{\pi/8}$ the latter acquires LRO. In analogy with the celebrated Haldane phase \cite{Haldane1983}, the LRO of this string correlation function signals the onset of interaction-induced symmetry-protected topological (SPT) regimes~\cite{Pollmann2012,Tang2015,Senthil2015,Shapourian2017,Montorsi2017}. 
Furthermore, in the SPT regime a fractional spin accumulates at the edge of the system. This can be seen by integrating the smooth component of the spin operator, given by $S^z_j \sim \frac{d_x}{2\pi}\partial_x\phi_s(x)$, at the edge of the system~\cite{Montorsi2017}. The calculation gives $S_{\text{edge}}^z=\lim_{d_x\to 0} \frac{1}{\sqrt{2\pi}}\int_{x_e}^{x_e+d_x}dx \partial_x\phi_s(x) = \pm\frac{1}{4}$, where we take the limit $d_x\to 0$ to find the accumulated spin at the edge $x_e$. This implies that the SPT phases are characterized by the LRO of the string correlation function as well as the appearance of protected edge modes. The latter also admits a direct quasiparticle interpretation. Specifically, the pinning of the bosonic field implies a gapped bulk, while the boundary enforces a solitonic interpolation between symmetry-related minima. This domain wall behaves as an emergent edge quasiparticle carrying a fractional spin quantum number, in close analogy with fractionalized excitations discussed in topological soliton theories of one-dimensional systems~\cite{Jackiw1976swf, Su1979sip}.
\begin{figure*}[t]
    \centering
    \includegraphics[width=2\columnwidth]{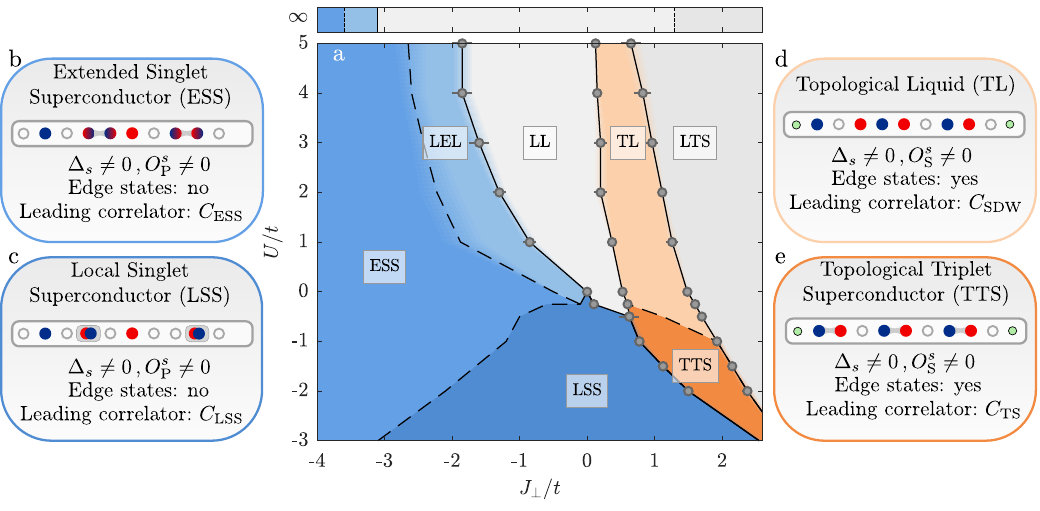}
    \caption{\textbf{Topology and superconductivity in a $t$\textnormal{--}$J$ model for magnetic atoms}. (a) Phase diagram of the $t$\textnormal{--}$J$ model Hamiltonian $H_{t\text{--}J}$ with
    total magnetization $S^z_\text{tot}=0$, density $\bar{n} = 1/2$, and $L=240$, as a function of $J_\perp/t$ and $U/t$. White and gray areas correspond to gapless regimes with $\Delta_s/t = 0$, while the colored regions indicate a finite spin gap, $\Delta_s/t \neq 0$, accompanied by long-range order in either the parity $O_\text{P}^s(r)$ (blue) or string $O_\text{S}^s(r)$ (orange) correlation functions. Solid black lines indicate phase transitions, while dashed lines denote crossovers.
    (b-e) Schematic representations of the topological and superconducting phases, highlighting their key characteristics. The full phase diagram reveals seven distinct phases: Extended Singlet Superconductor (ESS), Local Singlet Superconductor (LSS), Luther-Emery Liquid (LEL), Luttinger Liquid (LL), Topological Liquid (TL), Luttinger Triplet Superconductor (LTS), and Topological Triplet Superconductor (TTS). For $|J_\perp|\gg t,|U|$ (not shown) we find a regime of phase separation where empty and occupied sites are totally demixed.}
    \label{fig:Phase diagram Jp - U}
\end{figure*}

In addition to the behavior of the correlation functions in Eqs.\,\eqref{parity} and \eqref{SOP}, further important information can be derived by the Luttinger parameters $K_{c,s}$. Here, the condition $K_c > 1$ represents indeed a strict criterion~\cite{Giamarchi2003qpi} indicating the appearance of a one-dimensional superconducting phase.
Based on this discussion, our analysis of the Hamiltonian in Eq.\,\eqref{Hamiltonian bosonized spin} shows that for repulsive interactions $U > 0$, and $J_\perp < -U$ the field $\phi_s$ pins at $0$, thus implying the appearance of fermionic pairing captured by the LRO of $O_\text{P}^s(r)$. It has been shown that for open boundary conditions, this regime presents zero energy solutions that reflect the presence of edge bound states~\cite{Fabrizio1995iod}.
A similar behavior also takes place when the conditions $J_\perp < -U/2$ and $U<0$ are simultaneously fulfilled. This regime turns out to be particularly interesting, as the corresponding values of $J_\perp$ and $U$ further imply that the Luttinger parameter satisfies $K_c > 1$. As a consequence we expect one-dimensional superconductivity to occur.
In contrast, when $U > 0$ and $J_\perp > 0$ the bosonization analysis reveals that $\phi_s = \pm\sqrt{\frac{\pi}{8}}$, indicating the emergence of a SPT phase. 
Interestingly, the conditions $U < 0$ and $J_\perp > -U/2$ also support the emergence of a regime characterized by topological order. 
Crucially, for this range of parameters, we also find that $K_c > 1$. As a result, our low-energy analysis suggests the appearance of an important scenario: the coexistence of  topological order and superconductivity.
Finally, for all other combinations of $U$ and $J_\perp$ we find that $\phi_s$ is unpinned, and thus a gapless Luttinger liquid characterizes the ground state of Eq.\,\eqref{Hamiltonian bosonized spin}.

\subsection*{Strongly interacting topological and superconducting phases}

The preceding analysis reveals that the perturbative low-energy regime of Eq. \eqref{model} hosts a variety of intriguing many-body phases. In this section, we go beyond these findings by employing a Density-Matrix-Renormalization-Group (DMRG) \cite{White1992,SCHOLLWOCK201196} analysis. As shown in Fig.\,\ref{fig:Phase diagram Jp - U}(a), this numerical approach enables us to accurately explore the full phase diagram of $H_{t\text{--}J}$, overcoming the limitations of bosonization and revealing the structure of the ground state across a broader range of parameters, including strongly interacting non-perturbative regimes.\\
\begin{figure}[h]
    \centering
    \includegraphics{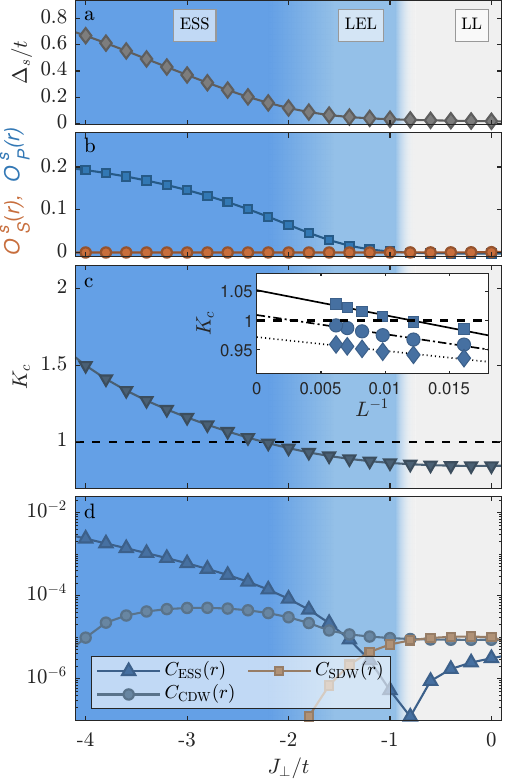}
    \caption{\textbf{Characterization of the Extended Singlet Superconductor (ESS) phase}. (a) Opening of the spin gap $\Delta_s/t$ for a negative spin-flip coupling $J_\perp$ induced by an attractive dipolar interaction. (b) Asymptotic value of the parity and string correlator $O_\text{P}^s(r)$ (blue squares), $O_\text{S}^s(r)$ (red circles). (c) The Luttinger parameter $K_c$. The inset shows $K_c$ plotted as a function of the inverse length $L^{-1}$, for three values of $J_\perp/t$ near the ESS to LEL transition (diamonds for $J_\perp/t = -2.0$, circles for $J_\perp/t = -2.2$ and squares for $J_\perp/t = -2.4$). The length $L$ in the inset represents the chain length after excluding 78 boundary sites (39 from each side) to minimize edge effects. (d) Asymptotic value of the spin density wave (SDW), charge density wave (CDW) and extended singlet superconductor (ESS) correlation functions (see main text). In order to minimize boundary effects, the quantities in (b-d) are calculated over the central $r=162$ sites. In all panels we set $U/t = 1.0$, density $\bar{n} = 1/2$ and the chain length $L = 240$.}
    \label{Panel 1: SS, LEL}
\end{figure}
We begin our analysis by fixing a repulsive onsite interaction and gradually decreasing $J_\perp$ from zero to negative values. Along this trajectory, Fig.\,\ref{Panel 1: SS, LEL}(a) shows that for $J_\perp/t$ large enough the spin gap $\Delta_s$ becomes slowly finite, signaling a Kosterlitz-Thouless transition from a gapless Luttinger liquid (LL) to a gapped regime. As revealed in Fig.\,\ref{Panel 1: SS, LEL}(b), the onset of a finite $\Delta_s$ is accompanied by the emergence of LRO of the parity correlation function $O_\text{P}^s(r)$. To further characterize this range of parameters, we compute the charge structure factor $S(q) = \frac{1}{L} \sum_{i,j} e^{\imath q(i-j)} \left(\langle n_i n_j \rangle - \langle n_i \rangle \langle n_j \rangle\right)$, where $n_i = n_{i,1/2} + n_{i,-1/2}$. From its low-momentum behavior, we extract the Luttinger parameter using the relation \cite{Giamarchi2003qpi}
\begin{align}
K_c = \lim_{q \to 0} \frac{\pi}{q} S(q).
\end{align}
As shown in Fig.\,\ref{Panel 1: SS, LEL}(c), decreasing $J_\perp$ further drives the system into a regime where $K_c > 1$, indicating the possible emergence of dominant superconducting correlations. We reinforce this by performing a finite-size scaling analysis of the Luttinger parameter $K_c$ around the transition point. 
This extrapolation to the thermodynamic limit (TDL) not only corroborates the transition to a superconducting phase, but also shows that the finite-size results at length 
$L = 240$ closely approximate the TDL.
As this prediction is only completely reliable in the low energy limit, we now explicitly examine the decay of relevant correlation functions. In particular, in one-dimensional systems the leading order of a quantum phase is determined by the correlation function that decays most slowly--or equivalently, maintains the largest value at long distances, see Methods. As shown in Fig.\,\ref{Panel 1: SS, LEL}(d), in the regime where $\Delta_s=0$, the spin ordering is dominant. This behavior is indeed unveiled by the largest asymptotic value of the spin density wave (SDW) correlator
\begin{align}\label{SDW}
C_\text{SDW}(r) = 4\left|\langle S_i^z S_{i+r}^z \rangle - \langle S_i^z \rangle \langle S_{i+r}^z \rangle\right|.
\end{align}
For intermediate negative values of $J_\perp/t$ giving rise to a finite spin gap, the dominant ordering becomes the charge one. This is evidenced by the charge density wave (CDW) correlation function,
\begin{align}\label{CDW}
C_\text{CDW}(r) =\left| \langle n_i n_{i+r} \rangle - \langle n_i \rangle \langle n_{i+r} \rangle\right|,
\end{align}
which exhibits the highest value at large $r$. Such a  behavior allows us to identify this regime as a Luther-Emery Liquid (LEL)\cite{Luther1974}, characterized by a finite spin gap and dominant CDW correlations. Notably, by further decreasing $J_\perp/t$ Fig.\,\ref{Panel 1: SS, LEL}(d) indicates that the charge ordering is replaced by an emergent extended singlet superconductor (ESS) phase captured by the slow decay of
\begin{align}\label{ESS}
C_\text{ESS}(r)= \left|\langle O_\text{ESS}^\dagger(i) O_\text{ESS}(i+r)\rangle\right|\,,
\end{align}
with $O_\text{ESS}^\dagger(i) = (1/\sqrt{2}) (c_{i,1/2}^\dagger c_{i+1,-1/2}^\dagger - c_{i,-1/2}^\dagger c_{i+1,1/2}^\dagger)$. Interestingly, the specific structure of $C_\text{ESS}(r)$ reveals that, in the ESS phase, the singlets are formed by bound pairs occupying two adjacent sites whose effective large tunneling amplitude gives rise to an interesting example of one-dimensional superconductivity, see Fig.\,\ref{fig:Phase diagram Jp - U}(b). Importantly, as indicated in Fig.\,\ref{fig:Phase diagram Jp - U}(a), the ESS phase spans a wide range of $U/t$ values, including both repulsive and attractive interactions. This broad stability range points to a high degree of experimental flexibility in realizing this superconducting phase.

\begin{figure}
    \centering
    \includegraphics[width=\columnwidth]{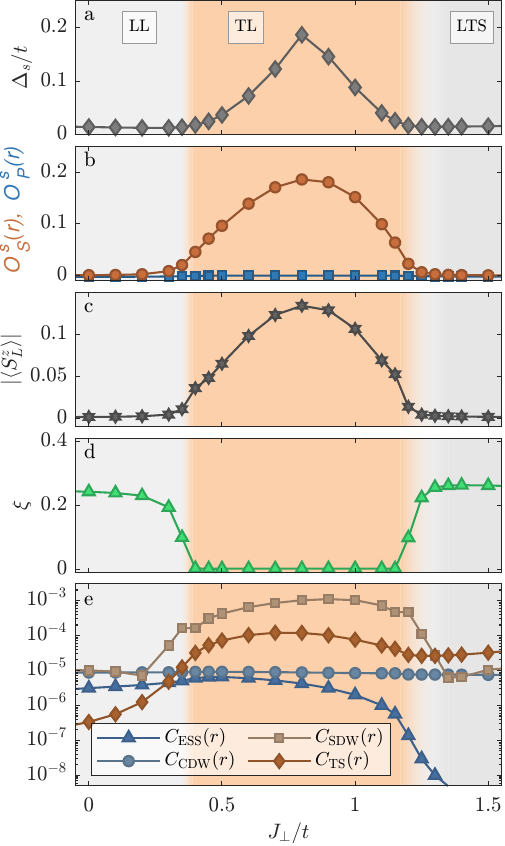}
    \caption{\textbf{Emergence of the topological liquid}. (a) Opening of the spin gap $\Delta_s/t$ for a positive spin-flip coupling $J_\perp$ induced by a repulsive dipolar interaction. (b) Asymptotic values of the parity and string correlators $O_\text{P}^s(r)$ (blue squares), $O_\text{S}^s(r)$ (red circles). (c) Edge magnetization $|\langle S_L^z \rangle|$. (d) Measure of the entanglement spectrum degeneracy $\xi$, where $\xi = 0$ implies topological order (see main text). (e) Asymptotic value of the spin density wave (SDW), charge density wave (CDW), extended singlet superconducting (ESS) and triplet superconducting (TS) correlation functions.
In all panels we set $U/t = 1.0$, density $\bar{n} = 1/2$ and chain length $L = 240$. In order to minimize boundary effects, the correlation functions in (b) and (e) are calculated over the central $r=162$ sites.}
    \label{Panel2: HL}
\end{figure}
We now turn our attention to the case where both $U/t$ and $J_\perp/t$ are positive. As shown in Fig.\,\ref{Panel2: HL}(a), the system remains in a gapless Luttinger liquid (LL) phase for weak spin-flip interactions, as indicated by the vanishing spin gap, $\Delta_s = 0$. However, for larger values of $J_\perp/t$, a phase transition occurs, marked by the emergence of a finite spin gap. In contrast to the previously discussed case, we find that the parity correlator $O_\text{P}^s(r)$ vanishes in this regime, while the string correlator $O_\text{S}^s(r)$ exhibits LRO, see Fig.\,\ref{Panel2: HL}(b). As discussed earlier, this behavior is a key signature of the emergence of symmetry-protected topological order. A well-known distinctive feature of topological phases is the appearance of edge states, which in a spin-gapped system manifest as finite edge magnetization $\langle S^z_1 \rangle = -\langle S^z_L \rangle \neq 0$ \cite{Fazzini2019iif, Montorsi2017}. This prediction is clearly confirmed by the results shown in Fig.\,\ref{Panel2: HL}(c). As a final hallmark of the appearance of topological order, we consider the entanglement spectrum (ES) whose even degeneracy provides a rigorous diagnostic of the SPT nature of a quantum phase \cite{Li2008,Pollmann2010,Turner2011}. Specifically, we compute the reduced density matrix $\rho_A=\sum_{n,N}\lambda_n^N\rho_n^N$ where $A=L/2$ is our considered system bi-partition and $\rho_n^N$ describes a pure state with $N$ particles with the corresponding eigenvalues $\lambda_n^N$ being the ES. In Fig~\ref{Panel2: HL}(d) we show that even ES degeneracy signaled by a vanishing $\xi = \lambda^N_1 - \lambda^N_2 + \lambda^N_3 - \lambda^N_4$ occurs exclusively in the region where the string order parameter $O_\text{S}^s(r)$ has LRO and the edge magnetization is finite. This unambiguously confirms the presence of an interaction induced topological phase. Our DMRG analysis further shows that increasing $J_\perp/t$ beyond a critical value destabilizes the SPT phase: both the spin gap $\Delta_s/t$ and edge magnetization vanish, and the entanglement spectrum no longer exhibits even degeneracy, with $\xi \neq 0$. To gain more precise insight into the nature of these different regimes, we once again turn to the behavior of correlation functions. As shown in Fig.\,\ref{Panel2: HL}(e), both in the gapless Luttinger liquid (LL) and in the SPT phase, the SDW correlator $C_\text{SDW}(r)$ captures the dominant spin ordering. This observation allows us to classify the SPT regime as a topological liquid (TL), protected by particle-hole and time reversal symmetries \cite{Montorsi2017}. An illustration of this regime is provided in Fig.\,\ref{fig:Phase diagram Jp - U}(d). 
It is important to emphasize that our system offers a concrete platform for realizing such an unconventional topological regime--markedly distinct from the interacting  SPT insulators that have been experimentally observed in ultracold atomic systems \cite{Leseleuc2019,Sompet2022,Walter2023,su2025topological}. Special attention must also be given to the regime that emerges for large positive values of $J_\perp/t$. In this case, the spin gap remains closed ($\Delta_s = 0$), and consequently, both the string and parity correlators, $O_\text{S}^{s}(r)$ and $O_\text{P}^{s}(r)$, vanish. Nonetheless, as shown in Fig.\,\ref{Panel2: HL}(e), the correlation function that decays most slowly--hence dominating the long-distance behavior--is the triplet superconducting correlator,
\begin{align}\label{TS}
C_\text{TS}(r) = \left|\langle O_\text{TS}^\dagger(i) O_\text{TS}(i+r) \rangle\right|,
\end{align}
where the triplet pairing operator is defined as
$O_\text{TS}^\dagger(i) = \frac{1}{\sqrt{2}} \left( c_{i,1/2}^\dagger c_{i+1,-1/2}^\dagger + c_{i,-1/2}^\dagger c_{i+1,1/2}^\dagger \right).$
This regime thus constitutes a compelling example of a gapless Luttinger triplet superconductor (LTS), a phase previously studied in other variants of the $t$--$J$ model~\cite{Gorshkov2011,Fazzini2019iif}. Notably, as shown in Fig.\,\ref{fig:Phase diagram Jp - U}(a), the LTS phase occupies a substantial portion of the phase diagram. This highlights once more the potential of our proposed setup for exploring unconventional superconducting states of matter in ultracold atomic systems.
\begin{figure}
    \centering
    \includegraphics[width=\columnwidth]{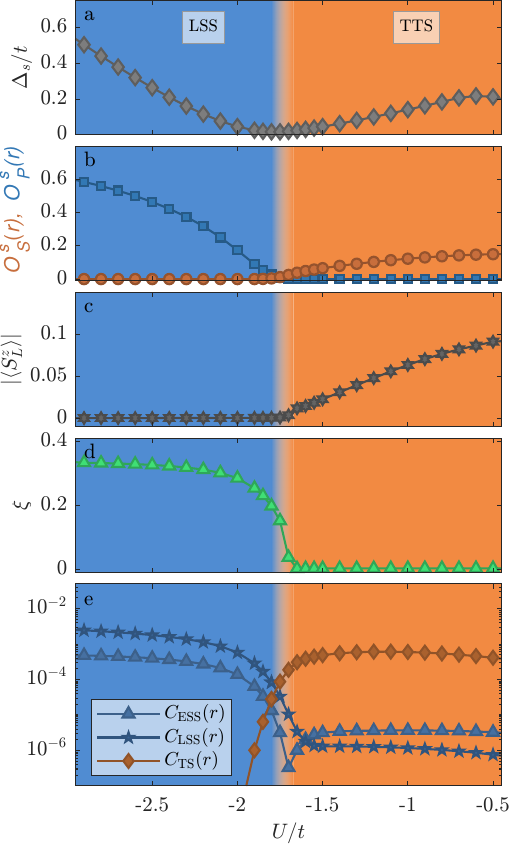}
    \caption{\textbf{Phase transition from local singlet to topological triplet superconductor phases}. (a) Spin gap $\Delta_s/t$ for attractive onsite interactions. (b) Asymptotic value of the parity and string correlator $O_\text{P}^s(r)$ (blue squares), $O_\text{S}^s(r)$ (red circles). (c) Edge magnetization $|\langle S_L^z\rangle|$. (d) Measure of the entanglement spectrum degeneracy $\xi$, where $\xi = 0$ implies topological order (see main text). (e) Asymptotic value of the extended singlet superconducting (ESS), local singlet superconducting (LSS) and triplet superconducting (TS) correlation functions. The parameters are fixed to $J_\perp/t = 1.25$, density $\bar{n} = 1/2$, and chain length $L = 240$. In order to minimize boundary effects, the correlation functions in (b) and (e) are calculated over the central $r=162$ sites.}
    \label{Panel 1: SS, HTS, LETS}
\end{figure}

\begin{figure}
    \centering
    \includegraphics[width=\columnwidth]{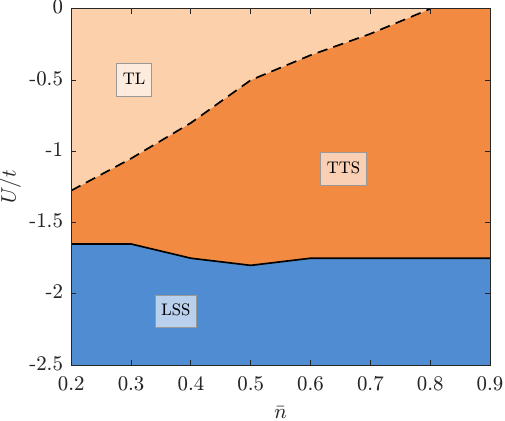}
    \caption{\textbf{Persistence of topological phases across densities.} Phase diagram of Eq.\,\eqref{model} at fixed $J_\perp/t = 1.25$ as a function of $U/t$ and particle density $\bar{n}$, for a chain of length $L = 240$. Colors and phase labels match those in Fig.\,\ref{fig:Phase diagram Jp - U}(a). Solid black lines indicate phase transitions, while dashed lines denote crossovers.}
    \label{fig:density_phase_diag}
\end{figure}

The versatility of our model allows us to explore even more intriguing states of matter by accessing the regime of positive $J_\perp/t$ in combination with attractive onsite interactions. In particular, by varying $U/t < 0$, Fig.\,\ref{Panel 1: SS, HTS, LETS}(a) reveals a phase transition, marked by the closing of the spin gap at a specific value of the onsite interaction. As shown in Fig.\,\ref{Panel 1: SS, HTS, LETS}(b), for strongly attractive $U/t$, the finite spin gap is associated with the LRO of the parity correlator $O_\text{P}^s(r)$. In contrast, as the onsite interaction becomes less negative, the string correlator $O_\text{S}^s(r)$ acquires LRO, signaling a qualitative change in the underlying phase. In this latter regime, the emergence of SPT order is confirmed by two key features: the appearance of finite edge magnetization, $\langle S^z_1\rangle = -\langle S^z_L\rangle \neq 0$, shown in Fig.\,\ref{Panel 1: SS, HTS, LETS}(c), and the even degeneracy of the entanglement spectrum, $\xi = 0$, shown in Fig.\,\ref{Panel 1: SS, HTS, LETS}(d). Additional insight into the $U/t < 0$ regime is provided by the behavior of the correlation function
\begin{align}\label{LS}
C_\text{LSS}(r) = \left|\langle O_\text{LSS}^\dagger(i) O_\text{LSS}(i+r) \rangle\right|,
\end{align}
with $O_\text{LSS}^\dagger(i) = c_{i,1/2}^\dagger c_{i,-1/2}^\dagger$. As shown in Fig.\,\ref{Panel 1: SS, HTS, LETS}(e), this correlator becomes dominant in the regime where $O_\text{P}^s(r)$ has LRO, namely for strongly attractive interactions. In this phase, singlet pairs are formed by fermions occupying the same lattice site, a feature made possible by the large attractive $U/t$, and characteristic of a local singlet superconductor (LSS). Importantly, this mechanism of onsite pairing--enabled by the combination of positive $J_\perp/t$ and negative $U/t$--is a unique feature of the $t$--$J$ model introduced in Eq.\,\eqref{model}. The relevance of our results becomes even more striking upon examining the SPT phase found at $U/t < 0$. As indicated in Fig.\,\ref{Panel 1: SS, HTS, LETS}(e), the string order coexists with dominant triplet superconducting correlations, $C_\text{TS}(r)$, which exhibit the slowest decay. This coexistence unambiguously points to the realization of a topological triplet superconductor (TTS), a phase where superconductivity and topology are intertwined. While similar phases have been predicted in models with competing magnetic couplings\cite{Fazzini2019iif,Montorsi2020}, here we uncover a novel mechanism: the emergence of TTS driven by strong spin-flip interactions combined with a small, but finite, density of doubly occupied sites. Finally, Fig.\,\ref{fig:density_phase_diag} demonstrates that the TTS phase remains robust across a wide range of $U/t$ and particle densities $\bar{n}$, providing significant flexibility for experimental realizations. We conclude this section by emphasizing a key point: as shown in Fig.\,\ref{fig:Phase diagram Jp - U}(a), the topological phases--TL and TTS--are absent when $U/t \to \infty$. This observation underscores once more the crucial role played by tunable onsite interactions, which, in our setup, can be varied independently of the anisotropic spin-spin coupling strength.
\subsection*{State preparation and detection scheme}

Based on the previous analysis, it is clear how $H_{t\text{--}J}$ can be realized and how such a versatile Hamiltonian can pave the way toward a deeper understanding of some of the most fascinating quantum phenomena in many-body states of matter--namely, topology, superconductivity, and topological superconductivity. However, two additional aspects, crucial for achieving these goals, must be carefully considered. The first is how to prepare these exotic phases; the second, how to observe them.

To achieve the first objective, we propose to use spin-polarized fermions in their lowest hyperfine state $|F,-F\rangle$ confined to a single \((x, y)\) layer of an anisotropic three-dimensional optical lattice. Within this layer, the  unit cell is rectangular, characterized by lattice spacings \(d_x\) and \(d_y\), where \(d_x < d_y\). To constrain the system to one-dimensional dynamics along the \(x\)-axis, the lattice depth in the \(y\)-direction is increased until tunneling along this axis becomes negligible. In practice we intend to have \(d_y\) $>$ 2\(d_x\) in order to suppress the magnetic interactions between different rows. Using an adiabatic loading of the atoms into the lattice, we expect to reach a low-entropy one-dimensional band insulator along $x$. The system size $L$ can be set by projecting hard walls onto the 1D systems via blue-detuned light sheets~\cite{Guardado2020sah}. For erbium and dysprosium, a blue-detuned repulsive potential can be reached using laser light in the UV region $<400\,$nm. 
While engineering the desired density and spin population, we plan to freeze the atom dynamics also along $x$ by increasing the corresponding lattice depth.
The target density of $\bar{n} \sim 0.5$ can be achieved by selectively removing atoms every second lattice site with a resonant light pulse~\cite{Gross2021}. For this we will implement local addressing, realized using local AC-Stark shifts induced by either a superlattice~\cite{Chalopin2025}, light-patterns formed by a spatial light modulator~\cite{Guardado2020sah}, or a movable, tightly focused laser beam~\cite{Weitenberg2011ssa}.
For spin manipulation, we will use a protocol recently demonstrated which employs a sequence of Rabi-pulse pairs coupling the ground state to an excited clock-type transition at $1299\,$nm, see Fig.\,\ref{fig1:illustration of system}(a). With this scheme we will prepare $|F,\pm1/2\rangle$ with resolved spin population.
At this stage, the on-site interaction $U$ and the spin-spin coupling $J_\perp$ can be set to their desired values by tuning the magnetic field magnitude and orientation~\cite{Baier2016ebh,Patscheider2020cde}. Finally, tunneling dynamics will be restored by adiabatically lowering the lattice depth along $x$, allowing the system to relax close to the ground states described above.

For our second goal, namely the detection of the various predicted phases, we plan to utilize well-established techniques from quantum gas microscopy~\cite{Gross2021}. Single-site imaging in our system with a lattice spacing of $d_x=266\,$nm along the main axis will be accomplished by using a high numerical-aperture objective~\cite{Sohmen2023asi} capable of resolving the lattice structure. Possible enhancements of detection fidelity can be reached by employing deep learning methods for the image analysis~\cite{Picard2019dla, Impertro2023aud} or magnification techniques~\cite{Asteria2021qgm,Su2023dqs}.
The spin at each lattice site can then be resolved via a shelving technique that uses the long-lived excited electronic state accessed through the clock-like transition. Specifically, by first pumping the $|F,+1/2\rangle$ atoms into the excited state, the remaining $|F,-1/2\rangle$ atoms can be detected using ultrafast imaging~\cite{Su2025hrf,su2025topological}. Afterwards, the excited-state atoms are pumped back into the $|F,-1/2\rangle$ ground state, and a second image is taken. This allows for the reconstruction of both spin and density at each lattice site in a single experimental run. 
With this scheme, accurate measurements of the local density $n_i$ and local magnetization $S^z_i$ become possible. This is of crucial importance: measurements of $n_i$ is used to extract quantities such as $K_c$, providing access to superconducting phases, while local magnetization measurements allow one to detect edge states (e.g., $S^z_1$ and $S^z_L$) and evaluate the string correlation function~\cite{Boll2016sad}. In addition, the use of a superlattice in combination with global addressing techniques allows for the measurement of pairing correlators \cite{impertro2024lra,mark2025emd}, thereby revealing the emergence of topological superconducting phases. Note that there might be also ways to directly measure the spin gap~\cite{Strinati2018sgs} or the entanglement spectrum~\cite{pichler2016mpf}.

\section*{Discussion}
We have presented a theoretical protocol tailored to a realistic experimental platform, opening a new avenue for the exploration of superconducting, topological, and topological superconducting phases. 
Specifically, leveraging the properties of the magnetic lanthanides erbium and dysprosium, we have derived a new version of the $t$--$J$ Hamiltonian. This model is characterized by an unprecedented level of versatility, with independently controllable hopping, onsite interaction, and magnetic couplings. This feature allows overcoming most of the limitations characteristic of $t$--$J$ Hamiltonians implemented with alkaline atoms~\cite{Hirthe2023,Bourgund2025,Chalopin2025} and their recently realized analogues with dipolar systems~\cite{carroll2024,douglas2024}.
In particular, our scheme gives access to regimes in which formation of double occupancies is allowed and strong anisotropic spin-spin interactions explicitly break the spin-rotational symmetry. 

Our theoretical analysis demonstrated that the aforementioned key aspects enable the emergence of fascinating many-body phases. Using both analytical and numerical methods, we have indeed shown that the ground state of our derived model can host, among the others, different superconducting states, a topological liquid, and a topological superconductor. In this regard, it is important to underline some crucial points: $1)$ The search for superconductivity in usual 2D Fermi-Hubbard models demands temperatures \cite{Bo-Xiao2017,Wietek2021} which are currently out of reach for ultracold atomic quantum systems. On the other hand, our setup just needs temperatures below the spin gap, i.e.~of the order of $T/t\lesssim0.2$, 
which are instead well within reach of the current experimental platforms using advanced entropy control schemes, with latest results achieving $T/t\approx0.05$~\cite{Xu2025};
$2)$ Our results unveiled a scheme to realize and probe symmetry-protected-topological phases where the particle motion is allowed, thus providing a concrete scheme to go beyond the many-body interaction induced symmetry-protected-topological insulators, experimentally investigated in ultracold atomic systems \cite{Leseleuc2019,Sompet2022,Walter2023,su2025topological}; $3)$ Compared to the paradigmatic example of one-dimensional topological superconductivity appearing in the celebrated Kitaev chain \cite{Kitaev_2001}, the topological-triplet-superconductor presents significative aspects of novelty. The most spectacular one is certainly the fact that such phase is completely induced by the presence of competing interacting processes, i.e.~$U$ and $J_\perp$. Furthermore, here we do not have any evidence that the edge states are Majorana fermions. Importantly, we also point out that in topological-triplet-superconductivity the number of particles is strictly conserved, thus, in principle, making its experimental realization less challenging. It is also worth stressing that we derived a realistic state preparation and detection protocol to realize and probe all the many-body phases we unveiled. We also highlight the fact that our scheme turns out to be very general as it directly applies without substantial differences to both Er and Dy magnetic atoms. As a consequence, our results represent a concrete and important step toward a deeper understanding of the intriguing states of matter emerging in strongly interacting fermionic quantum matter.

\section*{Methods}

\subsection*{Details on the calculations of the Hamiltonian parameters}

For the experimental values of $U$, $J_\perp$ and $t$ reported in Table \ref{tab:params}, we first numerically calculate the 3D Wannier functions of the lowest band for a given cubic lattice $\phi(\textbf{r})=\phi^x(x)\phi^y(y)\phi^z(z)$ with $(d_x,d_y,d_z)=(266,532,532)\,$nm and then evaluate the following terms numerically ($i$ and $j$ denote 
lattice sites along $x$)
\begin{equation}
t = -\int \text{d}^3\textbf{r}\, \phi_i^*(\textbf{r})\left(-\frac{\hbar^2\nabla^2}{2m}+V_\text{trap}(\textbf{r})\right)\phi_j(\textbf{r})\,,
\end{equation}
\begin{equation}
U = \frac{4\pi\hbar a_s}{m}\int \text{d}^3\textbf{r}|\phi_i(\textbf{r})|^4 \,,\\
\end{equation}
and
\begin{equation}\label{Vij}
V_{i,j} = \int \text{d}^3\textbf{r}\,\text{d}^3\textbf{r}'|\phi_i(\textbf{r})|^2V_\text{dd}(\textbf{r},\textbf{r}')|\phi_j(\textbf{r}')|^2\,,
\end{equation}
where 
\begin{equation}
V_\text{dd}(\textbf{r},\textbf{r}') = \frac{\mu_0\mu_{\rm B}^2g_F^2}{4\pi}\frac{1-3\cos^2\theta}{|\textbf{r}-\textbf{r}'|^3}\,,
\end{equation}
$\mu_{\rm B}$ is Bohr's magneton, $g_F$ is the Land\'e-factor and $\cos^2\theta = \langle \textbf{r}-\textbf{r}'|\vec{B}\rangle^2/(|r-r'||\vec{B}|)$. $\vec{B}$ is an external magnetic field that polarizes the dipoles. 
In this work, we consider spin-spin couplings that decay as the inverse cube of the distance.
The coupling strength is given by
\begin{equation}\label{Jperp}
    J_\perp = \frac{1}{4}\gamma \frac{V}{|i-j|^3}\,,
\end{equation}
where $V = V_{i, i+1}$ represents the nearest-neighbor interaction strength calculated through the integral of Eq.\,\ref{Vij}, and $\gamma$ is a prefactor determined by the quantum numbers of the system. Specifically, $\gamma$ is given by 
\begin{equation}
\gamma = \sqrt{F(F+1)-m_F(m_F-1)}\sqrt{F(F+1)-m_F'(m_F'+1)}\,
\end{equation}
with $F$ being the total angular momentum quantum number ($F=F_\text{Er}$ for erbium or $F=F_\text{Dy}$ for dysprosium), and $m_F=+1/2$ and $m_F'=-1/2$ representing the magnetic sublevels. 
While this expression assumes a $1/|i-j|^3$ scaling, the three-dimensional nature of the Wannier functions introduces slight deviations from this ideal behavior. To assess the accuracy of our approximation, we compute the interaction strength $V_{i,j}$ for up to three neighboring sites ($|i-j|_{\text{max}} = 3$). By comparing the approximate $J_\perp$ with the exact calculation, we find that the relative error remains below $4\%$ for the Dy parameters listed in Table \ref{tab:params}. This small error confirms the validity of the approximation and its applicability to the system we study. Note that Er and Dy have the same scaling with distance, and thus we expect our approximation to hold also for Er.

The on-site contact interaction $U$ and tunneling $t$ are evaluated via numerical integration on a grid with $1001\times1001\times1001$ points per unit cell of the lattice. We additionally checked that the on-site contribution of the DDI is negligible for typical parameters. The numerical integration of $V$ is done on a grid with $16\times33\times33$ points per unit cell of the lattice. The singularity in $V_\text{dd}$ is simply omitted, leading to a slight underestimation of the integral on the few percent level.

\subsection*{Bosonization analysis}
In this section, we provide additional details on the bosonization analysis, and for further technical aspects, we refer the reader to the following references  \cite{Gogolin1998,Giamarchi2003qpi,Senechal2004}. The general aim of this analysis is to derive an effective quantum field theory from a given Hamiltonian, and use that to study its ground state properties in the weak coupling limit. 
The first step is to rewrite the microscopic Hamiltonian in its continuum limit. For that we consider the low-energy regimes which are achieved for energies near the Fermi-points $\pm k_F$. This allows us to linearize the spectrum around these points and replace the annihilation operator with fermionic fields as
\begin{equation}\label{transformation c to R,L}
    c_{j,\alpha} \to \sqrt{d_x} [e^{i k_F x} \Psi_{R\sigma}(x) + e^{-i k_F x} \Psi_{L\sigma}(x)] \,, 
\end{equation}
where $\Psi_{R\sigma}(x)/\Psi_{L\sigma}(x)$ describe right- and left-moving particles and $x = j d_x$ ($d_x$ being the lattice spacing). We also substitute sums with integrals with the following prescription: $d_x \sum_j \to \int dx$. In the second step, we bosonize the newly derived Hamiltonian according to the standard bosonization-dictionary for fermionic bilinears. By following this procedure we map the microscopic Hamiltonian in Eq.\,\eqref{model} to the Sine-Gordon (SG) model in Eq.\,\eqref{Hamiltonian bosonized spin} \cite{Japaridze2000tsi, Dziurzik2004tsv}. This mapping allows us to understand the ground-state phase diagram of the original model in Eq.\,\eqref{model} by studying the low-energy properties of the newly derived SG model. This latter is described by the following set of renormalization group equations for the effective couplings $K_s(l)$ and $g_s(l)$
\begin{align}
    &\frac{dK_s(l)}{dl} = -\frac{1}{2}g_s^2(l)\,,\\
    &\frac{dg_s(l)}{dl} = 2(1-K_s(l))g_s(l)\,,
\end{align}
where $l = \ln{(d_x)}$.
Using the flow diagram of this set of equations we find that, whereas the quadratic terms in Eq.\,\eqref{Hamiltonian bosonized spin} favor unpinned bosonic fields $\phi_{c,s}$ and $\theta_{c,s}$, the massive term $\propto g_s$ wants the fields to be locked at the extrema of the cosine. Furthermore, we find that for $|g_s| > -2(1 - K_s)$ the system flows towards the strong coupling limit, which entails that the theory becomes massive, thus characterized by a finite spin gap, and that the term $\propto g_s$ becomes dominant with $\phi_s(x)$ pinned to the possible values $0, \pm \sqrt{\frac{\pi}{8}}$. The pinning to one or the other possible values is given by the sign of the mass term $g_s$. If $g_s > 0$ the field pins to $\phi_s(x) = \sqrt{\pi/8}$, whereas if $g_s < 0$ the pinning is $\phi_s = 0$. By inserting the definitions of $K_{c,s}$ and $g_s$ in the following conditions, we are able to find the regions of the parameter space in which the spin gap is open and the corresponding pinning value of the field $\phi_s$. This in turn allows us to associate a parity and string order parameter \cite{Montorsi2012}
\begin{align}
    &P^s \sim \langle\cos(\sqrt{2\pi}\phi_s)\rangle \quad\\
    &S^s \sim \langle\sin(\sqrt{2\pi}\phi_s)\rangle\,.
\end{align}
respectively. By directly inserting the pinning values of $\phi_s$, it is clear that $\phi_s = 0$ is uniquely associated with a finite parity order parameter $P^s$ and, consequently long-range order of $C_\text{P}^s(r)$. On the other hand, if $\phi_s = \pm \sqrt{\pi/8}$ the string order parameter $S^s$ is finite and therefore $C_\text{S}^s(r)$ displays long-range order. Finally, the regions in which $K_c > 1$, are proven to have dominant superconducting correlations \cite{Giamarchi2003qpi}. Thus, we use this condition to determine whether a specific region of the parameter space is characterized by superconducting order.

We can further characterize the Bosonization phase diagram by determining the dominant correlation function for each sector. To this end we compute the relevant correlators $\langle O^\dagger(x=0)O(x=r)\rangle$, with the order parameters given by
\begin{align}
    O^\dagger_\text{SDW} \sim &\cos({\sqrt{2\pi}\phi_c(x)+2k_Fx)\sin(\sqrt{2\pi}\phi_s(x))}\,,\\
    O^\dagger_\text{CDW} \sim &\sin(\sqrt{2\pi}\phi_c(x) - 2k_Fx)\cos(\sqrt{2\pi}\phi_s(x))\,,\\
    O_\text{LSS}^\dagger \sim &O_\text{ESS}^\dagger \sim \cos(\sqrt{2\pi}\phi_s(x))e^{i\sqrt{2\pi}\theta_c(x)} +\\
     &\sin(\sqrt{2\pi}\phi_c(x) + 2k_Fx)e^{i\sqrt{2\pi}\theta_c(x)}\,,\nonumber\\
    O_\text{TS}^\dagger \sim &\sin(\sqrt{2\pi}\phi_s(x))e^{i\sqrt{2\pi}\theta_c(x)} +\\
    &\sin(\sqrt{2\pi}\phi_c(x) + 2k_Fx)e^{i\sqrt{2\pi}\theta_c(x)}\,.\nonumber
\end{align}
Similarly as for the parity \eqref{POP} and string \eqref{SOP} correlation functions, depending on the pinning of the field $\phi_s$ we can identify power-law decays for different correlators. In particular, in the two sectors with $\phi_s$ pinned to $\pm \sqrt{\pi/8}$, we find a power-law decay for the spin density wave correlator $C_\text{SDW} = \langle O^\dagger_\text{SDW}(0) O_\text{SDW}(r)\rangle \sim r^{-K_c} \cos(\pi\bar{n}r)$ and the triplet-superconducting correlator $C_\text{TS} = \langle O^\dagger_\text{TS}(0) O_\text{TS}(r)\rangle \sim r^{-1/K_c}$. This means that the sector with $K_c < 1$ is characterized by dominant spin density wave order, whereas the sector $K_c > 1$ by triplet-superconducting instabilities. On the other hand, when $\phi_s$ is pinned to $0$, the charge density wave as well as local singlet superconducting and extended singlet superconducting correlators show a power-law decay. Thus, we identify dominant charge density wave order for $K_c < 1$ and singlet superconducting instabilities for $K_c > 1$. Finally, in the gapless regime we observe a power-law decay for the spin- and charge density wave correlators. More specifically, the calculation gives $C_\text{SDW} = C_\text{CDW} \sim r^{-\tilde{K}_s - K_c}$, with $\tilde{K_s} = 1 + \frac{1}{2\pi v_F}\sqrt{-4J_\perp (J_\perp + U)}$. Since in this regime $J_\perp>-U$ and $U>0$, we find $\tilde{K_s} > 1$.

\begin{figure}
    \centering
    \includegraphics{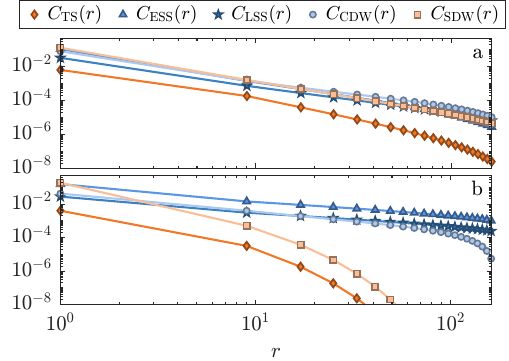}
    \caption{\textbf{Correlators for the phases in Fig.\,\ref{Panel 1: SS, LEL}}. (a) LEL phase where $J_\perp/t= -1.2$ and (b) ESS phase where $J_\perp/t = -3.4$. In (a-b) we fix $U/t = 1.0$, chain length $L = 240$, and density $\bar{n} = 1/2$. In order to minimize boundary effects, the correlation functions in (a) and (b) are calculated over the central $r=162$ sites. For better visualization of the decay of the correlation functions we plot every eighth value.}
    \label{Panel 1 SupMat: SS, LEL}
\end{figure}

\subsection*{Details on the DMRG analysis}

For the DMRG simulations we use the TeNPy package \cite{Hauschild2018Ens} and open boundary conditions with bond dimension up to $\chi = 800$. This ensures a maximal error in the ground state energy of the system of $\Delta E_\text{err} = 10^{-10}$. We perform the simulation for chains of length $L = 240$ and check that this size allows us to correctly approximate the thermodynamic limit. When necessary, we apply a weak local magnetic field to the edges of the chain to lift possible ground-state degeneracy. We consider the long-range spin-spin interaction up to the third nearest-neighbor. 

\subsection*{Decay of correlation functions}
\begin{figure}
    \centering
    \includegraphics{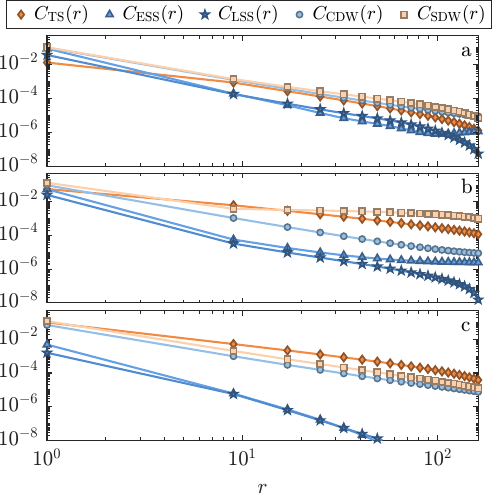}
    \caption{\textbf{Correlators for the phases in Fig.\,\ref{Panel2: HL}}. (a) LL phase where $J_\perp/t= 0.2$, (b) TL phase where $J_\perp/t = 0.8$ and (c) LTS phase where $J_\perp/t= 1.6$. In (a-c) we fix $U/t = 1.0$, chain length $L = 240$, and density $\bar{n} = 1/2$. In order to minimize boundary effects, the correlation functions in (a-c) are calculated over the central $r=162$ sites. For better visualization of the decay of the correlation functions we plot every eighth value.}
    \label{Panel 2 SupMat: LL, TL, LTS}
\end{figure}

In Figs.\,\ref{Panel 1: SS, LEL}, \ref{Panel2: HL} and \ref{Panel 1: SS, HTS, LETS}, we have characterized different phases by reporting the asymptotic value of the relevant correlation functions. In the following we show the decay of the correlation functions for each of the phases appearing in the phase diagram of Fig.\,\ref{fig:Phase diagram Jp - U}(a).

Fig.\,\ref{Panel 1 SupMat: SS, LEL} shows the correlation functions in (a) the Luther-Emery Liquid (LEL) and (b) Extended Singlet Superconductor (ESS) phases for $J_\perp/t = -1.2$ and $J_\perp/t = -3.4$ respectively. We fix the onsite interaction $U/t = 1.0$ and density $\bar{n} = 1/2$. We see that the dominant correlators are the charge density wave $C_\text{CDW}(r)$ and extended singlet superconductor $C_\text{ESS}(r)$, respectively.
In Fig.\,\ref{Panel 2 SupMat: LL, TL, LTS} we show the decay of the correlators in (a) the Luttinger Liquid (LL) phase with $J_\perp/t = 0.2$, (b) the Topological Liquid (TL) with $J_\perp/t = 0.8$ and (c) the Luttinger Triplet Superconductor (LTS) with $J_\perp/t = 1.6$. In all three subplots (a-c) we fix $U/t = 1.0$ and $\bar{n} = 1/2$. Here, we observe that both in the LL and TL phases the dominating correlator is the spin density wave (SDW), whereas in the LTS phase the triplet superconducting correlator is dominant.
Finally, in Fig.\,\ref{Panel 3 SupMat: LSS, HTS} we show the decay in (a) the Local Singlet Superconductor (LSS) with $U/t = -2.8$ and (b) the Topological Triplet Superconductor (TTS) with $U/t = -1.5$. We fix the spin-spin interaction $J_\perp/t = 1.25$, and density $\bar{n} = 1/2$. In the former, we observe  dominant local singlet superconducting whereas in the latter triplet superconducting correlations.
\begin{figure}[h]
    \centering
    \includegraphics{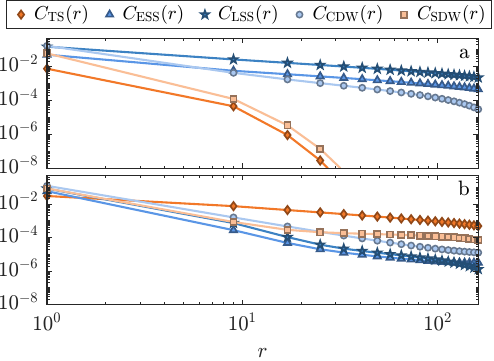}
    \caption{\textbf{Correlators for the phases in Fig.\,\ref{Panel 1: SS, HTS, LETS}}. (a) LSS phase where $U/t = -2.8$ and (b) TTS phase where $U/t = -1.5$. In (a-b) we fix $J_\perp/t = 1.25$, chain length $L = 240$, and density $\bar{n} = 1/2$. In order to minimize boundary effects, the correlation functions in (a) and (b) are calculated over the central $r=162$ sites. For better visualization of the decay of the correlation functions we plot every eighth value.}
    \label{Panel 3 SupMat: LSS, HTS}
\end{figure}

\newpage
\section*{Data Availability}
Data pertaining to this work are available via Zenodo \url{https://doi.org/10.5281/zenodo.18854391}.

\section*{Code Availability}
The DMRG simulations are performed using the publicly available TeNPy package~\cite{Hauschild2018Ens}.

\newpage

\section*{Acknowledgments}
We acknowledge A. Montorsi, E. Poli, A. M. Rey, L. Santos, and G. Valtolina for discussions. We also thank the other members of the Dipolar Quantum Gases group at the University of Innsbruck for useful discussions. This work was supported by the European Research Council through the Advanced Grant DyMETEr (\href{https://doi.org/10.3030/101054500}{10.3030/101054500}), the Austrian Science Fund (FWF) through the Cluster of Excellence QuantA (\href{https://doi.org/10.55776/COE1}{10.55776/COE1}). This research was also funded in part by the Austrian Science Fund (FWF) Grant (\href{https://doi.org/10.55776/PAT1597224}{10.55776/PAT1597224}). L. L. acknowledges funding from the Austrian Science Fund (FWF) within the DK-ALM (\href{https://doi.org/10.55776/W1259}{10.55776/W1259}).
T.B. thanks the Knut and Alice Wallenberg Foundation (GrantNo. KAW 2018.0217) and the Swedish Research Council (Grant No. 2022-03654 vr). L.B. acknowledges funding from the Italian MUR (PRIN DiQut Grant No. 2022523NA7). L.B. warmly acknowledges the Institute of Experimental Physics of the Innsbruck University for hospitality during the conceiving of this project.

\section*{Author Contributions}
F. F., M. J. M., and L. B. conceived the project; L. B. G. performed the analytical and DMRG calculations, with input from L. B.; F. F. and M. J. M. developed the model implementation; L. L., F. F., and M. J. M. carried out the state preparation and detection scheme; L. B. G., T. B., L. L., F. F.,  M. J. M. and L. B. contributed to the interpretation of the results and the writing of the manuscript.
\section*{Competing Interests}
The authors declare no competing interests.

\newpage 


\begin{thebibliography}{87}%
\makeatletter
\providecommand \@ifxundefined [1]{%
 \@ifx{#1\undefined}
}%
\providecommand \@ifnum [1]{%
 \ifnum #1\expandafter \@firstoftwo
 \else \expandafter \@secondoftwo
 \fi
}%
\providecommand \@ifx [1]{%
 \ifx #1\expandafter \@firstoftwo
 \else \expandafter \@secondoftwo
 \fi
}%
\providecommand \natexlab [1]{#1}%
\providecommand \enquote  [1]{``#1''}%
\providecommand \bibnamefont  [1]{#1}%
\providecommand \bibfnamefont [1]{#1}%
\providecommand \citenamefont [1]{#1}%
\providecommand \href@noop [0]{\@secondoftwo}%
\providecommand \href [0]{\begingroup \@sanitize@url \@href}%
\providecommand \@href[1]{\@@startlink{#1}\@@href}%
\providecommand \@@href[1]{\endgroup#1\@@endlink}%
\providecommand \@sanitize@url [0]{\catcode `\\12\catcode `\$12\catcode `\&12\catcode `\#12\catcode `\^12\catcode `\_12\catcode `\%12\relax}%
\providecommand \@@startlink[1]{}%
\providecommand \@@endlink[0]{}%
\providecommand \url  [0]{\begingroup\@sanitize@url \@url }%
\providecommand \@url [1]{\endgroup\@href {#1}{\urlprefix }}%
\providecommand \urlprefix  [0]{URL }%
\providecommand \Eprint [0]{\href }%
\providecommand \doibase [0]{https://doi.org/}%
\providecommand \selectlanguage [0]{\@gobble}%
\providecommand \bibinfo  [0]{\@secondoftwo}%
\providecommand \bibfield  [0]{\@secondoftwo}%
\providecommand \translation [1]{[#1]}%
\providecommand \BibitemOpen [0]{}%
\providecommand \bibitemStop [0]{}%
\providecommand \bibitemNoStop [0]{.\EOS\space}%
\providecommand \EOS [0]{\spacefactor3000\relax}%
\providecommand \BibitemShut  [1]{\csname bibitem#1\endcsname}%
\let\auto@bib@innerbib\@empty
%</preamble>
\bibitem [{\citenamefont {Auerbach}(1994)}]{assa}%
  \BibitemOpen
  \bibfield  {author} {\bibinfo {author} {\bibfnamefont {A.}~\bibnamefont {Auerbach}},\ }\href@noop {} {\emph {\bibinfo {title} {Interacting Electrons and Quantum Magnetism}}}\ (\bibinfo  {publisher} {Springer-Verlag},\ \bibinfo {year} {1994})\BibitemShut {NoStop}%
\bibitem [{\citenamefont {Sachdev}(2008)}]{Sachdev2008}%
  \BibitemOpen
  \bibfield  {author} {\bibinfo {author} {\bibfnamefont {S.}~\bibnamefont {Sachdev}},\ }\bibfield  {title} {\bibinfo {title} {Quantum magnetism and criticality},\ }\href {https://doi.org/10.1038/nphys894} {\bibfield  {journal} {\bibinfo  {journal} {Nature Physics}\ }\textbf {\bibinfo {volume} {4}},\ \bibinfo {pages} {173} (\bibinfo {year} {2008})}\BibitemShut {NoStop}%
\bibitem [{\citenamefont {Dagotto}(1994)}]{dagotto1994}%
  \BibitemOpen
  \bibfield  {author} {\bibinfo {author} {\bibfnamefont {E.}~\bibnamefont {Dagotto}},\ }\bibfield  {title} {\bibinfo {title} {Correlated electrons in high-temperature superconductors},\ }\href {https://doi.org/10.1103/RevModPhys.66.763} {\bibfield  {journal} {\bibinfo  {journal} {Rev. Mod. Phys.}\ }\textbf {\bibinfo {volume} {66}},\ \bibinfo {pages} {763} (\bibinfo {year} {1994})}\BibitemShut {NoStop}%
\bibitem [{\citenamefont {Lee}\ \emph {et~al.}(2006)\citenamefont {Lee}, \citenamefont {Nagaosa},\ and\ \citenamefont {Wen}}]{lee2006}%
  \BibitemOpen
  \bibfield  {author} {\bibinfo {author} {\bibfnamefont {P.~A.}\ \bibnamefont {Lee}}, \bibinfo {author} {\bibfnamefont {N.}~\bibnamefont {Nagaosa}},\ and\ \bibinfo {author} {\bibfnamefont {X.-G.}\ \bibnamefont {Wen}},\ }\bibfield  {title} {\bibinfo {title} {Doping a {M}ott insulator: Physics of high-temperature superconductivity},\ }\href {https://doi.org/10.1103/RevModPhys.78.17} {\bibfield  {journal} {\bibinfo  {journal} {Rev. Mod. Phys.}\ }\textbf {\bibinfo {volume} {78}},\ \bibinfo {pages} {17} (\bibinfo {year} {2006})}\BibitemShut {NoStop}%
\bibitem [{\citenamefont {Leijnse}\ and\ \citenamefont {Flensberg}(2012)}]{Leijnse_2012}%
  \BibitemOpen
  \bibfield  {author} {\bibinfo {author} {\bibfnamefont {M.}~\bibnamefont {Leijnse}}\ and\ \bibinfo {author} {\bibfnamefont {K.}~\bibnamefont {Flensberg}},\ }\bibfield  {title} {\bibinfo {title} {Introduction to topological superconductivity and {M}ajorana fermions},\ }\href {https://doi.org/10.1088/0268-1242/27/12/124003} {\bibfield  {journal} {\bibinfo  {journal} {Semiconductor Science and Technology}\ }\textbf {\bibinfo {volume} {27}},\ \bibinfo {pages} {124003} (\bibinfo {year} {2012})}\BibitemShut {NoStop}%
\bibitem [{\citenamefont {Sato}\ and\ \citenamefont {Fujimoto}(2016)}]{Sato2016}%
  \BibitemOpen
  \bibfield  {author} {\bibinfo {author} {\bibfnamefont {M.}~\bibnamefont {Sato}}\ and\ \bibinfo {author} {\bibfnamefont {S.}~\bibnamefont {Fujimoto}},\ }\bibfield  {title} {\bibinfo {title} {{M}ajorana fermions and topology in superconductors},\ }\href {https://doi.org/10.7566/JPSJ.85.072001} {\bibfield  {journal} {\bibinfo  {journal} {Journal of the Physical Society of Japan}\ }\textbf {\bibinfo {volume} {85}},\ \bibinfo {pages} {072001} (\bibinfo {year} {2016})}\BibitemShut {NoStop}%
\bibitem [{\citenamefont {Lewenstein}\ \emph {et~al.}(2012)\citenamefont {Lewenstein}, \citenamefont {Sanpera},\ and\ \citenamefont {Ahufinger}}]{Lewenstein2012}%
  \BibitemOpen
  \bibfield  {author} {\bibinfo {author} {\bibfnamefont {M.}~\bibnamefont {Lewenstein}}, \bibinfo {author} {\bibfnamefont {A.}~\bibnamefont {Sanpera}},\ and\ \bibinfo {author} {\bibfnamefont {V.}~\bibnamefont {Ahufinger}},\ }\href {https://books.google.it/books?id=WX4Xz7F6DdUC} {\emph {\bibinfo {title} {Ultracold Atoms in Optical Lattices: Simulating quantum many-body systems}}}\ (\bibinfo  {publisher} {OUP Oxford},\ \bibinfo {year} {2012})\BibitemShut {NoStop}%
\bibitem [{\citenamefont {Gross}\ and\ \citenamefont {Bloch}(2017)}]{Gross2017}%
  \BibitemOpen
  \bibfield  {author} {\bibinfo {author} {\bibfnamefont {C.}~\bibnamefont {Gross}}\ and\ \bibinfo {author} {\bibfnamefont {I.}~\bibnamefont {Bloch}},\ }\bibfield  {title} {\bibinfo {title} {Quantum simulations with ultracold atoms in optical lattices},\ }\href {https://doi.org/10.1126/science.aal3837} {\bibfield  {journal} {\bibinfo  {journal} {Science}\ }\textbf {\bibinfo {volume} {357}},\ \bibinfo {pages} {995} (\bibinfo {year} {2017})}\BibitemShut {NoStop}%
\bibitem [{\citenamefont {J{\"o}rdens}\ \emph {et~al.}(2008)\citenamefont {J{\"o}rdens}, \citenamefont {Strohmaier}, \citenamefont {G{\"u}nter}, \citenamefont {Moritz},\ and\ \citenamefont {Esslinger}}]{Jordens2008}%
  \BibitemOpen
  \bibfield  {author} {\bibinfo {author} {\bibfnamefont {R.}~\bibnamefont {J{\"o}rdens}}, \bibinfo {author} {\bibfnamefont {N.}~\bibnamefont {Strohmaier}}, \bibinfo {author} {\bibfnamefont {K.}~\bibnamefont {G{\"u}nter}}, \bibinfo {author} {\bibfnamefont {H.}~\bibnamefont {Moritz}},\ and\ \bibinfo {author} {\bibfnamefont {T.}~\bibnamefont {Esslinger}},\ }\bibfield  {title} {\bibinfo {title} {A {M}ott insulator of fermionic atoms in an optical lattice},\ }\href {https://doi.org/10.1038/nature07244} {\bibfield  {journal} {\bibinfo  {journal} {Nature}\ }\textbf {\bibinfo {volume} {455}},\ \bibinfo {pages} {204} (\bibinfo {year} {2008})}\BibitemShut {NoStop}%
\bibitem [{\citenamefont {Tusi}\ \emph {et~al.}(2022)\citenamefont {Tusi}, \citenamefont {Franchi}, \citenamefont {Livi}, \citenamefont {Baumann}, \citenamefont {Benedicto~Orenes}, \citenamefont {Del~Re}, \citenamefont {Barfknecht}, \citenamefont {Zhou}, \citenamefont {Inguscio}, \citenamefont {Cappellini}, \citenamefont {Capone}, \citenamefont {Catani},\ and\ \citenamefont {Fallani}}]{Tusi2022}%
  \BibitemOpen
  \bibfield  {author} {\bibinfo {author} {\bibfnamefont {D.}~\bibnamefont {Tusi}}, \bibinfo {author} {\bibfnamefont {L.}~\bibnamefont {Franchi}}, \bibinfo {author} {\bibfnamefont {L.~F.}\ \bibnamefont {Livi}}, \bibinfo {author} {\bibfnamefont {K.}~\bibnamefont {Baumann}}, \bibinfo {author} {\bibfnamefont {D.}~\bibnamefont {Benedicto~Orenes}}, \bibinfo {author} {\bibfnamefont {L.}~\bibnamefont {Del~Re}}, \bibinfo {author} {\bibfnamefont {R.~E.}\ \bibnamefont {Barfknecht}}, \bibinfo {author} {\bibfnamefont {T.~W.}\ \bibnamefont {Zhou}}, \bibinfo {author} {\bibfnamefont {M.}~\bibnamefont {Inguscio}}, \bibinfo {author} {\bibfnamefont {G.}~\bibnamefont {Cappellini}}, \bibinfo {author} {\bibfnamefont {M.}~\bibnamefont {Capone}}, \bibinfo {author} {\bibfnamefont {J.}~\bibnamefont {Catani}},\ and\ \bibinfo {author} {\bibfnamefont {L.}~\bibnamefont {Fallani}},\ }\bibfield  {title} {\bibinfo {title} {Flavour-selective localization in interacting lattice fermions},\ }\href {https://doi.org/10.1038/s41567-022-01726-5}
  {\bibfield  {journal} {\bibinfo  {journal} {Nature Physics}\ }\textbf {\bibinfo {volume} {18}},\ \bibinfo {pages} {1201} (\bibinfo {year} {2022})}\BibitemShut {NoStop}%
\bibitem [{\citenamefont {Mazurenko}\ \emph {et~al.}(2017)\citenamefont {Mazurenko}, \citenamefont {Chiu}, \citenamefont {Ji}, \citenamefont {Parsons}, \citenamefont {Kan{\'a}sz-Nagy}, \citenamefont {Schmidt}, \citenamefont {Grusdt}, \citenamefont {Demler}, \citenamefont {Greif},\ and\ \citenamefont {Greiner}}]{Mazurenko2017}%
  \BibitemOpen
  \bibfield  {author} {\bibinfo {author} {\bibfnamefont {A.}~\bibnamefont {Mazurenko}}, \bibinfo {author} {\bibfnamefont {C.~S.}\ \bibnamefont {Chiu}}, \bibinfo {author} {\bibfnamefont {G.}~\bibnamefont {Ji}}, \bibinfo {author} {\bibfnamefont {M.~F.}\ \bibnamefont {Parsons}}, \bibinfo {author} {\bibfnamefont {M.}~\bibnamefont {Kan{\'a}sz-Nagy}}, \bibinfo {author} {\bibfnamefont {R.}~\bibnamefont {Schmidt}}, \bibinfo {author} {\bibfnamefont {F.}~\bibnamefont {Grusdt}}, \bibinfo {author} {\bibfnamefont {E.}~\bibnamefont {Demler}}, \bibinfo {author} {\bibfnamefont {D.}~\bibnamefont {Greif}},\ and\ \bibinfo {author} {\bibfnamefont {M.}~\bibnamefont {Greiner}},\ }\bibfield  {title} {\bibinfo {title} {A cold-atom {F}ermi--{H}ubbard antiferromagnet},\ }\href {https://doi.org/10.1038/nature22362} {\bibfield  {journal} {\bibinfo  {journal} {Nature}\ }\textbf {\bibinfo {volume} {545}},\ \bibinfo {pages} {462} (\bibinfo {year} {2017})}\BibitemShut {NoStop}%
\bibitem [{\citenamefont {Shao}\ \emph {et~al.}(2024)\citenamefont {Shao}, \citenamefont {Wang}, \citenamefont {Zhu}, \citenamefont {Zhu}, \citenamefont {Sun}, \citenamefont {Chen}, \citenamefont {Zhang}, \citenamefont {Fan}, \citenamefont {Deng}, \citenamefont {Yao}, \citenamefont {Chen},\ and\ \citenamefont {Pan}}]{shao2024}%
  \BibitemOpen
  \bibfield  {author} {\bibinfo {author} {\bibfnamefont {H.-J.}\ \bibnamefont {Shao}}, \bibinfo {author} {\bibfnamefont {Y.-X.}\ \bibnamefont {Wang}}, \bibinfo {author} {\bibfnamefont {D.-Z.}\ \bibnamefont {Zhu}}, \bibinfo {author} {\bibfnamefont {Y.-S.}\ \bibnamefont {Zhu}}, \bibinfo {author} {\bibfnamefont {H.-N.}\ \bibnamefont {Sun}}, \bibinfo {author} {\bibfnamefont {S.-Y.}\ \bibnamefont {Chen}}, \bibinfo {author} {\bibfnamefont {C.}~\bibnamefont {Zhang}}, \bibinfo {author} {\bibfnamefont {Z.-J.}\ \bibnamefont {Fan}}, \bibinfo {author} {\bibfnamefont {Y.}~\bibnamefont {Deng}}, \bibinfo {author} {\bibfnamefont {X.-C.}\ \bibnamefont {Yao}}, \bibinfo {author} {\bibfnamefont {Y.-A.}\ \bibnamefont {Chen}},\ and\ \bibinfo {author} {\bibfnamefont {J.-W.}\ \bibnamefont {Pan}},\ }\href {https://arxiv.org/abs/2402.14605} {\bibinfo {title} {Observation of the antiferromagnetic phase transition in the fermionic {H}ubbard model}} (\bibinfo {year} {2024}),\ \Eprint {https://arxiv.org/abs/2402.14605} {arXiv:2402.14605
  [cond-mat.quant-gas]} \BibitemShut {NoStop}%
\bibitem [{\citenamefont {Lebrat}\ \emph {et~al.}(2024)\citenamefont {Lebrat}, \citenamefont {Kale}, \citenamefont {Kendrick}, \citenamefont {Xu}, \citenamefont {Gang}, \citenamefont {Nikolaenko}, \citenamefont {Sachdev},\ and\ \citenamefont {Greiner}}]{lebrat2024}%
  \BibitemOpen
  \bibfield  {author} {\bibinfo {author} {\bibfnamefont {M.}~\bibnamefont {Lebrat}}, \bibinfo {author} {\bibfnamefont {A.}~\bibnamefont {Kale}}, \bibinfo {author} {\bibfnamefont {L.~H.}\ \bibnamefont {Kendrick}}, \bibinfo {author} {\bibfnamefont {M.}~\bibnamefont {Xu}}, \bibinfo {author} {\bibfnamefont {Y.}~\bibnamefont {Gang}}, \bibinfo {author} {\bibfnamefont {A.}~\bibnamefont {Nikolaenko}}, \bibinfo {author} {\bibfnamefont {S.}~\bibnamefont {Sachdev}},\ and\ \bibinfo {author} {\bibfnamefont {M.}~\bibnamefont {Greiner}},\ }\href {https://arxiv.org/abs/2404.17555} {\bibinfo {title} {Ferrimagnetism of ultracold fermions in a multi-band {H}ubbard system}} (\bibinfo {year} {2024}),\ \Eprint {https://arxiv.org/abs/2404.17555} {arXiv:2404.17555 [cond-mat.quant-gas]} \BibitemShut {NoStop}%
\bibitem [{\citenamefont {Sompet}\ \emph {et~al.}(2022)\citenamefont {Sompet}, \citenamefont {Hirthe}, \citenamefont {Bourgund}, \citenamefont {Chalopin}, \citenamefont {Bibo}, \citenamefont {Koepsell}, \citenamefont {Bojovi{\'c}}, \citenamefont {Verresen}, \citenamefont {Pollmann}, \citenamefont {Salomon}, \citenamefont {Gross}, \citenamefont {Hilker},\ and\ \citenamefont {Bloch}}]{Sompet2022}%
  \BibitemOpen
  \bibfield  {author} {\bibinfo {author} {\bibfnamefont {P.}~\bibnamefont {Sompet}}, \bibinfo {author} {\bibfnamefont {S.}~\bibnamefont {Hirthe}}, \bibinfo {author} {\bibfnamefont {D.}~\bibnamefont {Bourgund}}, \bibinfo {author} {\bibfnamefont {T.}~\bibnamefont {Chalopin}}, \bibinfo {author} {\bibfnamefont {J.}~\bibnamefont {Bibo}}, \bibinfo {author} {\bibfnamefont {J.}~\bibnamefont {Koepsell}}, \bibinfo {author} {\bibfnamefont {P.}~\bibnamefont {Bojovi{\'c}}}, \bibinfo {author} {\bibfnamefont {R.}~\bibnamefont {Verresen}}, \bibinfo {author} {\bibfnamefont {F.}~\bibnamefont {Pollmann}}, \bibinfo {author} {\bibfnamefont {G.}~\bibnamefont {Salomon}}, \bibinfo {author} {\bibfnamefont {C.}~\bibnamefont {Gross}}, \bibinfo {author} {\bibfnamefont {T.~A.}\ \bibnamefont {Hilker}},\ and\ \bibinfo {author} {\bibfnamefont {I.}~\bibnamefont {Bloch}},\ }\bibfield  {title} {\bibinfo {title} {Realizing the symmetry-protected {H}aldane phase in {F}ermi--{H}ubbard ladders},\ }\href {https://doi.org/10.1038/s41586-022-04688-z}
  {\bibfield  {journal} {\bibinfo  {journal} {Nature}\ }\textbf {\bibinfo {volume} {606}},\ \bibinfo {pages} {484} (\bibinfo {year} {2022})}\BibitemShut {NoStop}%
\bibitem [{\citenamefont {Walter}\ \emph {et~al.}(2023)\citenamefont {Walter}, \citenamefont {Zhu}, \citenamefont {G{\"a}chter}, \citenamefont {Minguzzi}, \citenamefont {Roschinski}, \citenamefont {Sandholzer}, \citenamefont {Viebahn},\ and\ \citenamefont {Esslinger}}]{Walter2023}%
  \BibitemOpen
  \bibfield  {author} {\bibinfo {author} {\bibfnamefont {A.-S.}\ \bibnamefont {Walter}}, \bibinfo {author} {\bibfnamefont {Z.}~\bibnamefont {Zhu}}, \bibinfo {author} {\bibfnamefont {M.}~\bibnamefont {G{\"a}chter}}, \bibinfo {author} {\bibfnamefont {J.}~\bibnamefont {Minguzzi}}, \bibinfo {author} {\bibfnamefont {S.}~\bibnamefont {Roschinski}}, \bibinfo {author} {\bibfnamefont {K.}~\bibnamefont {Sandholzer}}, \bibinfo {author} {\bibfnamefont {K.}~\bibnamefont {Viebahn}},\ and\ \bibinfo {author} {\bibfnamefont {T.}~\bibnamefont {Esslinger}},\ }\bibfield  {title} {\bibinfo {title} {Quantization and its breakdown in a {H}ubbard--{T}houless pump},\ }\href {https://doi.org/10.1038/s41567-023-02145-w} {\bibfield  {journal} {\bibinfo  {journal} {Nature Physics}\ }\textbf {\bibinfo {volume} {19}},\ \bibinfo {pages} {1471} (\bibinfo {year} {2023})}\BibitemShut {NoStop}%
\bibitem [{\citenamefont {Brown}\ \emph {et~al.}(2019)\citenamefont {Brown}, \citenamefont {Mitra}, \citenamefont {Guardado-Sanchez}, \citenamefont {Nourafkan}, \citenamefont {Reymbaut}, \citenamefont {H{\'e}bert}, \citenamefont {Bergeron}, \citenamefont {Tremblay}, \citenamefont {Kokalj}, \citenamefont {Huse}, \citenamefont {Schau{\ss}},\ and\ \citenamefont {Bakr}}]{Brown2019}%
  \BibitemOpen
  \bibfield  {author} {\bibinfo {author} {\bibfnamefont {P.~T.}\ \bibnamefont {Brown}}, \bibinfo {author} {\bibfnamefont {D.}~\bibnamefont {Mitra}}, \bibinfo {author} {\bibfnamefont {E.}~\bibnamefont {Guardado-Sanchez}}, \bibinfo {author} {\bibfnamefont {R.}~\bibnamefont {Nourafkan}}, \bibinfo {author} {\bibfnamefont {A.}~\bibnamefont {Reymbaut}}, \bibinfo {author} {\bibfnamefont {C.-D.}\ \bibnamefont {H{\'e}bert}}, \bibinfo {author} {\bibfnamefont {S.}~\bibnamefont {Bergeron}}, \bibinfo {author} {\bibfnamefont {A.-M.~S.}\ \bibnamefont {Tremblay}}, \bibinfo {author} {\bibfnamefont {J.}~\bibnamefont {Kokalj}}, \bibinfo {author} {\bibfnamefont {D.~A.}\ \bibnamefont {Huse}}, \bibinfo {author} {\bibfnamefont {P.}~\bibnamefont {Schau{\ss}}},\ and\ \bibinfo {author} {\bibfnamefont {W.~S.}\ \bibnamefont {Bakr}},\ }\bibfield  {title} {\bibinfo {title} {Bad metallic transport in a cold atom {F}ermi-{H}ubbard system},\ }\href {https://doi.org/10.1126/science.aat4134} {\bibfield  {journal} {\bibinfo  {journal}
  {Science}\ }\textbf {\bibinfo {volume} {363}},\ \bibinfo {pages} {379} (\bibinfo {year} {2019})}\BibitemShut {NoStop}%
\bibitem [{\citenamefont {Hartke}\ \emph {et~al.}(2023)\citenamefont {Hartke}, \citenamefont {Oreg}, \citenamefont {Turnbaugh}, \citenamefont {Jia},\ and\ \citenamefont {Zwierlein}}]{Hartke2023}%
  \BibitemOpen
  \bibfield  {author} {\bibinfo {author} {\bibfnamefont {T.}~\bibnamefont {Hartke}}, \bibinfo {author} {\bibfnamefont {B.}~\bibnamefont {Oreg}}, \bibinfo {author} {\bibfnamefont {C.}~\bibnamefont {Turnbaugh}}, \bibinfo {author} {\bibfnamefont {N.}~\bibnamefont {Jia}},\ and\ \bibinfo {author} {\bibfnamefont {M.}~\bibnamefont {Zwierlein}},\ }\bibfield  {title} {\bibinfo {title} {Direct observation of nonlocal fermion pairing in an attractive {F}ermi-{H}ubbard gas},\ }\href {https://doi.org/10.1126/science.ade4245} {\bibfield  {journal} {\bibinfo  {journal} {Science}\ }\textbf {\bibinfo {volume} {381}},\ \bibinfo {pages} {82} (\bibinfo {year} {2023})}\BibitemShut {NoStop}%
\bibitem [{\citenamefont {Zhou}\ \emph {et~al.}(2023)\citenamefont {Zhou}, \citenamefont {Cappellini}, \citenamefont {Tusi}, \citenamefont {Franchi}, \citenamefont {Parravicini}, \citenamefont {Repellin}, \citenamefont {Greschner}, \citenamefont {Inguscio}, \citenamefont {Giamarchi}, \citenamefont {Filippone}, \citenamefont {Catani},\ and\ \citenamefont {Fallani}}]{Zhou2023}%
  \BibitemOpen
  \bibfield  {author} {\bibinfo {author} {\bibfnamefont {T.-W.}\ \bibnamefont {Zhou}}, \bibinfo {author} {\bibfnamefont {G.}~\bibnamefont {Cappellini}}, \bibinfo {author} {\bibfnamefont {D.}~\bibnamefont {Tusi}}, \bibinfo {author} {\bibfnamefont {L.}~\bibnamefont {Franchi}}, \bibinfo {author} {\bibfnamefont {J.}~\bibnamefont {Parravicini}}, \bibinfo {author} {\bibfnamefont {C.}~\bibnamefont {Repellin}}, \bibinfo {author} {\bibfnamefont {S.}~\bibnamefont {Greschner}}, \bibinfo {author} {\bibfnamefont {M.}~\bibnamefont {Inguscio}}, \bibinfo {author} {\bibfnamefont {T.}~\bibnamefont {Giamarchi}}, \bibinfo {author} {\bibfnamefont {M.}~\bibnamefont {Filippone}}, \bibinfo {author} {\bibfnamefont {J.}~\bibnamefont {Catani}},\ and\ \bibinfo {author} {\bibfnamefont {L.}~\bibnamefont {Fallani}},\ }\bibfield  {title} {\bibinfo {title} {Observation of universal hall response in strongly interacting fermions},\ }\href {https://doi.org/10.1126/science.add1969} {\bibfield  {journal} {\bibinfo  {journal} {Science}\ }\textbf
  {\bibinfo {volume} {381}},\ \bibinfo {pages} {427} (\bibinfo {year} {2023})}\BibitemShut {NoStop}%
\bibitem [{\citenamefont {Bourgund}\ \emph {et~al.}(2025)\citenamefont {Bourgund}, \citenamefont {Chalopin}, \citenamefont {Bojovi{\'c}}, \citenamefont {Schl{\"o}mer}, \citenamefont {Wang}, \citenamefont {Franz}, \citenamefont {Hirthe}, \citenamefont {Bohrdt}, \citenamefont {Grusdt}, \citenamefont {Bloch},\ and\ \citenamefont {Hilker}}]{Bourgund2025}%
  \BibitemOpen
  \bibfield  {author} {\bibinfo {author} {\bibfnamefont {D.}~\bibnamefont {Bourgund}}, \bibinfo {author} {\bibfnamefont {T.}~\bibnamefont {Chalopin}}, \bibinfo {author} {\bibfnamefont {P.}~\bibnamefont {Bojovi{\'c}}}, \bibinfo {author} {\bibfnamefont {H.}~\bibnamefont {Schl{\"o}mer}}, \bibinfo {author} {\bibfnamefont {S.}~\bibnamefont {Wang}}, \bibinfo {author} {\bibfnamefont {T.}~\bibnamefont {Franz}}, \bibinfo {author} {\bibfnamefont {S.}~\bibnamefont {Hirthe}}, \bibinfo {author} {\bibfnamefont {A.}~\bibnamefont {Bohrdt}}, \bibinfo {author} {\bibfnamefont {F.}~\bibnamefont {Grusdt}}, \bibinfo {author} {\bibfnamefont {I.}~\bibnamefont {Bloch}},\ and\ \bibinfo {author} {\bibfnamefont {T.~A.}\ \bibnamefont {Hilker}},\ }\bibfield  {title} {\bibinfo {title} {Formation of individual stripes in a mixed-dimensional cold-atom {F}ermi--{H}ubbard system},\ }\href {https://doi.org/10.1038/s41586-024-08270-7} {\bibfield  {journal} {\bibinfo  {journal} {Nature}\ }\textbf {\bibinfo {volume} {637}},\ \bibinfo {pages} {57}
  (\bibinfo {year} {2025})}\BibitemShut {NoStop}%
\bibitem [{\citenamefont {Chalopin}\ \emph {et~al.}(2026)\citenamefont {Chalopin}, \citenamefont {Bojović}, \citenamefont {Wang}, \citenamefont {Franz}, \citenamefont {Sinha}, \citenamefont {Wang}, \citenamefont {Bourgund}, \citenamefont {Obermeyer}, \citenamefont {Grusdt}, \citenamefont {Bohrdt}, \citenamefont {Pollet}, \citenamefont {Wietek}, \citenamefont {Georges}, \citenamefont {Hilker},\ and\ \citenamefont {Bloch}}]{chalopin2024}%
  \BibitemOpen
  \bibfield  {author} {\bibinfo {author} {\bibfnamefont {T.}~\bibnamefont {Chalopin}}, \bibinfo {author} {\bibfnamefont {P.}~\bibnamefont {Bojović}}, \bibinfo {author} {\bibfnamefont {S.}~\bibnamefont {Wang}}, \bibinfo {author} {\bibfnamefont {T.}~\bibnamefont {Franz}}, \bibinfo {author} {\bibfnamefont {A.}~\bibnamefont {Sinha}}, \bibinfo {author} {\bibfnamefont {Z.}~\bibnamefont {Wang}}, \bibinfo {author} {\bibfnamefont {D.}~\bibnamefont {Bourgund}}, \bibinfo {author} {\bibfnamefont {J.}~\bibnamefont {Obermeyer}}, \bibinfo {author} {\bibfnamefont {F.}~\bibnamefont {Grusdt}}, \bibinfo {author} {\bibfnamefont {A.}~\bibnamefont {Bohrdt}}, \bibinfo {author} {\bibfnamefont {L.}~\bibnamefont {Pollet}}, \bibinfo {author} {\bibfnamefont {A.}~\bibnamefont {Wietek}}, \bibinfo {author} {\bibfnamefont {A.}~\bibnamefont {Georges}}, \bibinfo {author} {\bibfnamefont {T.}~\bibnamefont {Hilker}},\ and\ \bibinfo {author} {\bibfnamefont {I.}~\bibnamefont {Bloch}},\ }\bibfield  {title} {\bibinfo {title} {Observation of
  emergent scaling of spin–charge correlations at the onset of the pseudogap},\ }\href {https://doi.org/10.1073/pnas.2525539123} {\bibfield  {journal} {\bibinfo  {journal} {Proceedings of the National Academy of Sciences}\ }\textbf {\bibinfo {volume} {123}},\ \bibinfo {pages} {e2525539123} (\bibinfo {year} {2026})}\BibitemShut {NoStop}%
\bibitem [{\citenamefont {{H}ubbard}\ and\ \citenamefont {Flowers}(1963)}]{Hubbard1963}%
  \BibitemOpen
  \bibfield  {author} {\bibinfo {author} {\bibfnamefont {J.}~\bibnamefont {{H}ubbard}}\ and\ \bibinfo {author} {\bibfnamefont {B.~H.}\ \bibnamefont {Flowers}},\ }\bibfield  {title} {\bibinfo {title} {Electron correlations in narrow energy bands},\ }\href {https://doi.org/10.1098/rspa.1963.0204} {\bibfield  {journal} {\bibinfo  {journal} {Proceedings of the Royal Society of London. Series A. Mathematical and Physical Sciences}\ }\textbf {\bibinfo {volume} {276}},\ \bibinfo {pages} {238} (\bibinfo {year} {1963})}\BibitemShut {NoStop}%
\bibitem [{\citenamefont {Chao}\ \emph {et~al.}(1978)\citenamefont {Chao}, \citenamefont {Spa\l{}ek},\ and\ \citenamefont {Ole\ifmmode~\acute{s}\else \'{s}\fi{}}}]{Chao1978}%
  \BibitemOpen
  \bibfield  {author} {\bibinfo {author} {\bibfnamefont {K.~A.}\ \bibnamefont {Chao}}, \bibinfo {author} {\bibfnamefont {J.}~\bibnamefont {Spa\l{}ek}},\ and\ \bibinfo {author} {\bibfnamefont {A.~M.}\ \bibnamefont {Ole\ifmmode~\acute{s}\else \'{s}\fi{}}},\ }\bibfield  {title} {\bibinfo {title} {Canonical perturbation expansion of the {H}ubbard model},\ }\href {https://doi.org/10.1103/PhysRevB.18.3453} {\bibfield  {journal} {\bibinfo  {journal} {Phys. Rev. B}\ }\textbf {\bibinfo {volume} {18}},\ \bibinfo {pages} {3453} (\bibinfo {year} {1978})}\BibitemShut {NoStop}%
\bibitem [{\citenamefont {Hirthe}\ \emph {et~al.}(2023)\citenamefont {Hirthe}, \citenamefont {Chalopin}, \citenamefont {Bourgund}, \citenamefont {Bojovi{\'c}}, \citenamefont {Bohrdt}, \citenamefont {Demler}, \citenamefont {Grusdt}, \citenamefont {Bloch},\ and\ \citenamefont {Hilker}}]{Hirthe2023}%
  \BibitemOpen
  \bibfield  {author} {\bibinfo {author} {\bibfnamefont {S.}~\bibnamefont {Hirthe}}, \bibinfo {author} {\bibfnamefont {T.}~\bibnamefont {Chalopin}}, \bibinfo {author} {\bibfnamefont {D.}~\bibnamefont {Bourgund}}, \bibinfo {author} {\bibfnamefont {P.}~\bibnamefont {Bojovi{\'c}}}, \bibinfo {author} {\bibfnamefont {A.}~\bibnamefont {Bohrdt}}, \bibinfo {author} {\bibfnamefont {E.}~\bibnamefont {Demler}}, \bibinfo {author} {\bibfnamefont {F.}~\bibnamefont {Grusdt}}, \bibinfo {author} {\bibfnamefont {I.}~\bibnamefont {Bloch}},\ and\ \bibinfo {author} {\bibfnamefont {T.~A.}\ \bibnamefont {Hilker}},\ }\bibfield  {title} {\bibinfo {title} {Magnetically mediated hole pairing in fermionic ladders of ultracold atoms},\ }\href {https://doi.org/10.1038/s41586-022-05437-y} {\bibfield  {journal} {\bibinfo  {journal} {Nature}\ }\textbf {\bibinfo {volume} {613}},\ \bibinfo {pages} {463} (\bibinfo {year} {2023})}\BibitemShut {NoStop}%
\bibitem [{\citenamefont {Chalopin}\ \emph {et~al.}(2025)\citenamefont {Chalopin}, \citenamefont {Bojovi\ifmmode~\acute{c}\else \'{c}\fi{}}, \citenamefont {Bourgund}, \citenamefont {Wang}, \citenamefont {Franz}, \citenamefont {Bloch},\ and\ \citenamefont {Hilker}}]{Chalopin2025}%
  \BibitemOpen
  \bibfield  {author} {\bibinfo {author} {\bibfnamefont {T.}~\bibnamefont {Chalopin}}, \bibinfo {author} {\bibfnamefont {P.}~\bibnamefont {Bojovi\ifmmode~\acute{c}\else \'{c}\fi{}}}, \bibinfo {author} {\bibfnamefont {D.}~\bibnamefont {Bourgund}}, \bibinfo {author} {\bibfnamefont {S.}~\bibnamefont {Wang}}, \bibinfo {author} {\bibfnamefont {T.}~\bibnamefont {Franz}}, \bibinfo {author} {\bibfnamefont {I.}~\bibnamefont {Bloch}},\ and\ \bibinfo {author} {\bibfnamefont {T.}~\bibnamefont {Hilker}},\ }\bibfield  {title} {\bibinfo {title} {Optical superlattice for engineering {H}ubbard couplings in quantum simulation},\ }\href {https://doi.org/10.1103/PhysRevLett.134.053402} {\bibfield  {journal} {\bibinfo  {journal} {Phys. Rev. Lett.}\ }\textbf {\bibinfo {volume} {134}},\ \bibinfo {pages} {053402} (\bibinfo {year} {2025})}\BibitemShut {NoStop}%
\bibitem [{\citenamefont {Gorshkov}\ \emph {et~al.}(2011)\citenamefont {Gorshkov}, \citenamefont {Manmana}, \citenamefont {Chen}, \citenamefont {Ye}, \citenamefont {Demler}, \citenamefont {Lukin},\ and\ \citenamefont {Rey}}]{Gorshkov2011}%
  \BibitemOpen
  \bibfield  {author} {\bibinfo {author} {\bibfnamefont {A.~V.}\ \bibnamefont {Gorshkov}}, \bibinfo {author} {\bibfnamefont {S.~R.}\ \bibnamefont {Manmana}}, \bibinfo {author} {\bibfnamefont {G.}~\bibnamefont {Chen}}, \bibinfo {author} {\bibfnamefont {J.}~\bibnamefont {Ye}}, \bibinfo {author} {\bibfnamefont {E.}~\bibnamefont {Demler}}, \bibinfo {author} {\bibfnamefont {M.~D.}\ \bibnamefont {Lukin}},\ and\ \bibinfo {author} {\bibfnamefont {A.~M.}\ \bibnamefont {Rey}},\ }\bibfield  {title} {\bibinfo {title} {Tunable superfluidity and quantum magnetism with ultracold polar molecules},\ }\href {https://doi.org/10.1103/PhysRevLett.107.115301} {\bibfield  {journal} {\bibinfo  {journal} {Phys. Rev. Lett.}\ }\textbf {\bibinfo {volume} {107}},\ \bibinfo {pages} {115301} (\bibinfo {year} {2011})}\BibitemShut {NoStop}%
\bibitem [{\citenamefont {Carroll}\ \emph {et~al.}(2025)\citenamefont {Carroll}, \citenamefont {Hirzler}, \citenamefont {Miller}, \citenamefont {Wellnitz}, \citenamefont {Muleady}, \citenamefont {Lin}, \citenamefont {Zamarski}, \citenamefont {Wang}, \citenamefont {Bohn}, \citenamefont {Rey},\ and\ \citenamefont {Ye}}]{carroll2024}%
  \BibitemOpen
  \bibfield  {author} {\bibinfo {author} {\bibfnamefont {A.~N.}\ \bibnamefont {Carroll}}, \bibinfo {author} {\bibfnamefont {H.}~\bibnamefont {Hirzler}}, \bibinfo {author} {\bibfnamefont {C.}~\bibnamefont {Miller}}, \bibinfo {author} {\bibfnamefont {D.}~\bibnamefont {Wellnitz}}, \bibinfo {author} {\bibfnamefont {S.~R.}\ \bibnamefont {Muleady}}, \bibinfo {author} {\bibfnamefont {J.}~\bibnamefont {Lin}}, \bibinfo {author} {\bibfnamefont {K.~P.}\ \bibnamefont {Zamarski}}, \bibinfo {author} {\bibfnamefont {R.~R.~W.}\ \bibnamefont {Wang}}, \bibinfo {author} {\bibfnamefont {J.~L.}\ \bibnamefont {Bohn}}, \bibinfo {author} {\bibfnamefont {A.~M.}\ \bibnamefont {Rey}},\ and\ \bibinfo {author} {\bibfnamefont {J.}~\bibnamefont {Ye}},\ }\bibfield  {title} {\bibinfo {title} {Observation of generalized t-j spin dynamics with tunable dipolar interactions},\ }\href {https://doi.org/10.1126/science.adq0911} {\bibfield  {journal} {\bibinfo  {journal} {Science}\ }\textbf {\bibinfo {volume} {388}},\ \bibinfo {pages} {381}
  (\bibinfo {year} {2025})}\BibitemShut {NoStop}%
\bibitem [{\citenamefont {Douglas}\ \emph {et~al.}(2025)\citenamefont {Douglas}, \citenamefont {Kaxiras}, \citenamefont {Su}, \citenamefont {Szurek}, \citenamefont {Singh}, \citenamefont {Markovi\ifmmode~\acute{c}\else \'{c}\fi{}},\ and\ \citenamefont {Greiner}}]{douglas2024}%
  \BibitemOpen
  \bibfield  {author} {\bibinfo {author} {\bibfnamefont {A.}~\bibnamefont {Douglas}}, \bibinfo {author} {\bibfnamefont {V.}~\bibnamefont {Kaxiras}}, \bibinfo {author} {\bibfnamefont {L.}~\bibnamefont {Su}}, \bibinfo {author} {\bibfnamefont {M.}~\bibnamefont {Szurek}}, \bibinfo {author} {\bibfnamefont {V.}~\bibnamefont {Singh}}, \bibinfo {author} {\bibfnamefont {O.}~\bibnamefont {Markovi\ifmmode~\acute{c}\else \'{c}\fi{}}},\ and\ \bibinfo {author} {\bibfnamefont {M.}~\bibnamefont {Greiner}},\ }\bibfield  {title} {\bibinfo {title} {Spin squeezing with itinerant magnetic dipoles},\ }\href {https://doi.org/10.1103/shj7-9kb3} {\bibfield  {journal} {\bibinfo  {journal} {Phys. Rev. X}\ }\textbf {\bibinfo {volume} {15}},\ \bibinfo {pages} {041021} (\bibinfo {year} {2025})}\BibitemShut {NoStop}%
\bibitem [{\citenamefont {Claude}\ \emph {et~al.}(2024)\citenamefont {Claude}, \citenamefont {Lafforgue}, \citenamefont {Houwman}, \citenamefont {Mark},\ and\ \citenamefont {Ferlaino}}]{Claude2024}%
  \BibitemOpen
  \bibfield  {author} {\bibinfo {author} {\bibfnamefont {F.}~\bibnamefont {Claude}}, \bibinfo {author} {\bibfnamefont {L.}~\bibnamefont {Lafforgue}}, \bibinfo {author} {\bibfnamefont {J.~J.~A.}\ \bibnamefont {Houwman}}, \bibinfo {author} {\bibfnamefont {M.~J.}\ \bibnamefont {Mark}},\ and\ \bibinfo {author} {\bibfnamefont {F.}~\bibnamefont {Ferlaino}},\ }\bibfield  {title} {\bibinfo {title} {Optical manipulation of spin states in ultracold magnetic atoms via an inner-shell {H}z transition},\ }\href {https://doi.org/10.1103/PhysRevResearch.6.L042016} {\bibfield  {journal} {\bibinfo  {journal} {Phys. Rev. Res.}\ }\textbf {\bibinfo {volume} {6}},\ \bibinfo {pages} {L042016} (\bibinfo {year} {2024})}\BibitemShut {NoStop}%
\bibitem [{\citenamefont {Chin}\ \emph {et~al.}(2010)\citenamefont {Chin}, \citenamefont {Grimm}, \citenamefont {Julienne},\ and\ \citenamefont {Tiesinga}}]{Chin2010fri}%
  \BibitemOpen
  \bibfield  {author} {\bibinfo {author} {\bibfnamefont {C.}~\bibnamefont {Chin}}, \bibinfo {author} {\bibfnamefont {R.}~\bibnamefont {Grimm}}, \bibinfo {author} {\bibfnamefont {P.}~\bibnamefont {Julienne}},\ and\ \bibinfo {author} {\bibfnamefont {E.}~\bibnamefont {Tiesinga}},\ }\bibfield  {title} {\bibinfo {title} {Feshbach resonances in ultracold gases},\ }\href {https://doi.org/10.1103/RevModPhys.82.1225} {\bibfield  {journal} {\bibinfo  {journal} {Rev. Mod. Phys.}\ }\textbf {\bibinfo {volume} {82}},\ \bibinfo {pages} {1225} (\bibinfo {year} {2010})}\BibitemShut {NoStop}%
\bibitem [{\citenamefont {Baier}\ \emph {et~al.}(2018)\citenamefont {Baier}, \citenamefont {Petter}, \citenamefont {Becher}, \citenamefont {Patscheider}, \citenamefont {Natale}, \citenamefont {Chomaz}, \citenamefont {Mark},\ and\ \citenamefont {Ferlaino}}]{Baier2018roa}%
  \BibitemOpen
  \bibfield  {author} {\bibinfo {author} {\bibfnamefont {S.}~\bibnamefont {Baier}}, \bibinfo {author} {\bibfnamefont {D.}~\bibnamefont {Petter}}, \bibinfo {author} {\bibfnamefont {J.~H.}\ \bibnamefont {Becher}}, \bibinfo {author} {\bibfnamefont {A.}~\bibnamefont {Patscheider}}, \bibinfo {author} {\bibfnamefont {G.}~\bibnamefont {Natale}}, \bibinfo {author} {\bibfnamefont {L.}~\bibnamefont {Chomaz}}, \bibinfo {author} {\bibfnamefont {M.~J.}\ \bibnamefont {Mark}},\ and\ \bibinfo {author} {\bibfnamefont {F.}~\bibnamefont {Ferlaino}},\ }\bibfield  {title} {\bibinfo {title} {Realization of a strongly interacting {F}ermi gas of dipolar atoms},\ }\href {https://doi.org/10.1103/PhysRevLett.121.093602} {\bibfield  {journal} {\bibinfo  {journal} {Phys. Rev. Lett.}\ }\textbf {\bibinfo {volume} {121}},\ \bibinfo {pages} {093602} (\bibinfo {year} {2018})}\BibitemShut {NoStop}%
\bibitem [{\citenamefont {Gross}\ and\ \citenamefont {Bakr}(2021)}]{Gross2021}%
  \BibitemOpen
  \bibfield  {author} {\bibinfo {author} {\bibfnamefont {C.}~\bibnamefont {Gross}}\ and\ \bibinfo {author} {\bibfnamefont {W.~S.}\ \bibnamefont {Bakr}},\ }\bibfield  {title} {\bibinfo {title} {Quantum gas microscopy for single atom and spin detection},\ }\href {https://doi.org/10.1038/s41567-021-01370-5} {\bibfield  {journal} {\bibinfo  {journal} {Nature Physics}\ }\textbf {\bibinfo {volume} {17}},\ \bibinfo {pages} {1316} (\bibinfo {year} {2021})}\BibitemShut {NoStop}%
\bibitem [{\citenamefont {Lauprêtre}\ \emph {et~al.}(2025)\citenamefont {Lauprêtre}, \citenamefont {Rey}, \citenamefont {Vernac},\ and\ \citenamefont {Laburthe-Tolra}}]{laupretre2025}%
  \BibitemOpen
  \bibfield  {author} {\bibinfo {author} {\bibfnamefont {T.}~\bibnamefont {Lauprêtre}}, \bibinfo {author} {\bibfnamefont {A.~M.}\ \bibnamefont {Rey}}, \bibinfo {author} {\bibfnamefont {L.}~\bibnamefont {Vernac}},\ and\ \bibinfo {author} {\bibfnamefont {B.}~\bibnamefont {Laburthe-Tolra}},\ }\href {https://arxiv.org/abs/2501.11402} {\bibinfo {title} {Probing coherences and itinerant magnetism in a dipolar lattice gas}} (\bibinfo {year} {2025}),\ \Eprint {https://arxiv.org/abs/2501.11402} {arXiv:2501.11402 [cond-mat.quant-gas]} \BibitemShut {NoStop}%
\bibitem [{\citenamefont {Manmana}\ \emph {et~al.}(2011)\citenamefont {Manmana}, \citenamefont {Hazzard}, \citenamefont {Chen}, \citenamefont {Feiguin},\ and\ \citenamefont {Rey}}]{Manmana2011SUN}%
  \BibitemOpen
  \bibfield  {author} {\bibinfo {author} {\bibfnamefont {S.~R.}\ \bibnamefont {Manmana}}, \bibinfo {author} {\bibfnamefont {K.~R.~A.}\ \bibnamefont {Hazzard}}, \bibinfo {author} {\bibfnamefont {G.}~\bibnamefont {Chen}}, \bibinfo {author} {\bibfnamefont {A.~E.}\ \bibnamefont {Feiguin}},\ and\ \bibinfo {author} {\bibfnamefont {A.~M.}\ \bibnamefont {Rey}},\ }\bibfield  {title} {\bibinfo {title} {Su$(n)$ magnetism in chains of ultracold alkaline-earth-metal atoms: Mott transitions and quantum correlations},\ }\href {https://doi.org/10.1103/PhysRevA.84.043601} {\bibfield  {journal} {\bibinfo  {journal} {Phys. Rev. A}\ }\textbf {\bibinfo {volume} {84}},\ \bibinfo {pages} {043601} (\bibinfo {year} {2011})}\BibitemShut {NoStop}%
\bibitem [{\citenamefont {Esslinger}(2010)}]{Esslinger2010}%
  \BibitemOpen
  \bibfield  {author} {\bibinfo {author} {\bibfnamefont {T.}~\bibnamefont {Esslinger}},\ }\bibfield  {title} {\bibinfo {title} {{F}ermi-{H}ubbard physics with atoms in an optical lattice},\ }\href {https://doi.org/https://doi.org/10.1146/annurev-conmatphys-070909-104059} {\bibfield  {journal} {\bibinfo  {journal} {Annual Review of Condensed Matter Physics}\ }\textbf {\bibinfo {volume} {1}},\ \bibinfo {pages} {129} (\bibinfo {year} {2010})}\BibitemShut {NoStop}%
\bibitem [{\citenamefont {Lecomte}\ \emph {et~al.}(2025)\citenamefont {Lecomte}, \citenamefont {Journeaux}, \citenamefont {Veschambre}, \citenamefont {Dalibard},\ and\ \citenamefont {Lopes}}]{Lecomte2025}%
  \BibitemOpen
  \bibfield  {author} {\bibinfo {author} {\bibfnamefont {M.}~\bibnamefont {Lecomte}}, \bibinfo {author} {\bibfnamefont {A.}~\bibnamefont {Journeaux}}, \bibinfo {author} {\bibfnamefont {J.}~\bibnamefont {Veschambre}}, \bibinfo {author} {\bibfnamefont {J.}~\bibnamefont {Dalibard}},\ and\ \bibinfo {author} {\bibfnamefont {R.}~\bibnamefont {Lopes}},\ }\bibfield  {title} {\bibinfo {title} {Production and stabilization of a spin mixture of ultracold dipolar bose gases},\ }\href {https://doi.org/10.1103/PhysRevLett.134.013402} {\bibfield  {journal} {\bibinfo  {journal} {Phys. Rev. Lett.}\ }\textbf {\bibinfo {volume} {134}},\ \bibinfo {pages} {013402} (\bibinfo {year} {2025})}\BibitemShut {NoStop}%
\bibitem [{\citenamefont {Patscheider}\ \emph {et~al.}(2020)\citenamefont {Patscheider}, \citenamefont {Zhu}, \citenamefont {Chomaz}, \citenamefont {Petter}, \citenamefont {Baier}, \citenamefont {Rey}, \citenamefont {Ferlaino},\ and\ \citenamefont {Mark}}]{Patscheider2020cde}%
  \BibitemOpen
  \bibfield  {author} {\bibinfo {author} {\bibfnamefont {A.}~\bibnamefont {Patscheider}}, \bibinfo {author} {\bibfnamefont {B.}~\bibnamefont {Zhu}}, \bibinfo {author} {\bibfnamefont {L.}~\bibnamefont {Chomaz}}, \bibinfo {author} {\bibfnamefont {D.}~\bibnamefont {Petter}}, \bibinfo {author} {\bibfnamefont {S.}~\bibnamefont {Baier}}, \bibinfo {author} {\bibfnamefont {A.-M.}\ \bibnamefont {Rey}}, \bibinfo {author} {\bibfnamefont {F.}~\bibnamefont {Ferlaino}},\ and\ \bibinfo {author} {\bibfnamefont {M.~J.}\ \bibnamefont {Mark}},\ }\bibfield  {title} {\bibinfo {title} {Controlling dipolar exchange interactions in a dense three-dimensional array of large-spin fermions},\ }\href {https://doi.org/10.1103/PhysRevResearch.2.023050} {\bibfield  {journal} {\bibinfo  {journal} {Phys. Rev. Res.}\ }\textbf {\bibinfo {volume} {2}},\ \bibinfo {pages} {023050} (\bibinfo {year} {2020})}\BibitemShut {NoStop}%
\bibitem [{\citenamefont {de~Paz}\ \emph {et~al.}(2013)\citenamefont {de~Paz}, \citenamefont {Sharma}, \citenamefont {Chotia}, \citenamefont {Mar\'echal}, \citenamefont {Huckans}, \citenamefont {Pedri}, \citenamefont {Santos}, \citenamefont {Gorceix}, \citenamefont {Vernac},\ and\ \citenamefont {Laburthe-Tolra}}]{dePaz2013nqm}%
  \BibitemOpen
  \bibfield  {author} {\bibinfo {author} {\bibfnamefont {A.}~\bibnamefont {de~Paz}}, \bibinfo {author} {\bibfnamefont {A.}~\bibnamefont {Sharma}}, \bibinfo {author} {\bibfnamefont {A.}~\bibnamefont {Chotia}}, \bibinfo {author} {\bibfnamefont {E.}~\bibnamefont {Mar\'echal}}, \bibinfo {author} {\bibfnamefont {J.~H.}\ \bibnamefont {Huckans}}, \bibinfo {author} {\bibfnamefont {P.}~\bibnamefont {Pedri}}, \bibinfo {author} {\bibfnamefont {L.}~\bibnamefont {Santos}}, \bibinfo {author} {\bibfnamefont {O.}~\bibnamefont {Gorceix}}, \bibinfo {author} {\bibfnamefont {L.}~\bibnamefont {Vernac}},\ and\ \bibinfo {author} {\bibfnamefont {B.}~\bibnamefont {Laburthe-Tolra}},\ }\bibfield  {title} {\bibinfo {title} {Nonequilibrium quantum magnetism in a dipolar lattice gas},\ }\href {https://doi.org/10.1103/PhysRevLett.111.185305} {\bibfield  {journal} {\bibinfo  {journal} {Phys. Rev. Lett.}\ }\textbf {\bibinfo {volume} {111}},\ \bibinfo {pages} {185305} (\bibinfo {year} {2013})}\BibitemShut {NoStop}%
\bibitem [{\citenamefont {Baier}\ \emph {et~al.}(2016)\citenamefont {Baier}, \citenamefont {Mark}, \citenamefont {Petter}, \citenamefont {Aikawa}, \citenamefont {Chomaz}, \citenamefont {Cai}, \citenamefont {Baranov}, \citenamefont {Zoller},\ and\ \citenamefont {Ferlaino}}]{Baier2016ebh}%
  \BibitemOpen
  \bibfield  {author} {\bibinfo {author} {\bibfnamefont {S.}~\bibnamefont {Baier}}, \bibinfo {author} {\bibfnamefont {M.~J.}\ \bibnamefont {Mark}}, \bibinfo {author} {\bibfnamefont {D.}~\bibnamefont {Petter}}, \bibinfo {author} {\bibfnamefont {K.}~\bibnamefont {Aikawa}}, \bibinfo {author} {\bibfnamefont {L.}~\bibnamefont {Chomaz}}, \bibinfo {author} {\bibfnamefont {Z.}~\bibnamefont {Cai}}, \bibinfo {author} {\bibfnamefont {M.}~\bibnamefont {Baranov}}, \bibinfo {author} {\bibfnamefont {P.}~\bibnamefont {Zoller}},\ and\ \bibinfo {author} {\bibfnamefont {F.}~\bibnamefont {Ferlaino}},\ }\bibfield  {title} {\bibinfo {title} {Extended {B}ose-{H}ubbard models with ultracold magnetic atoms},\ }\href {https://doi.org/10.1126/science.aac9812} {\bibfield  {journal} {\bibinfo  {journal} {Science}\ }\textbf {\bibinfo {volume} {352}},\ \bibinfo {pages} {201} (\bibinfo {year} {2016})}\BibitemShut {NoStop}%
\bibitem [{\citenamefont {Lepoutre}\ \emph {et~al.}(2019)\citenamefont {Lepoutre}, \citenamefont {Schachenmayer}, \citenamefont {Gabardos}, \citenamefont {Zhu}, \citenamefont {Naylor}, \citenamefont {Mar{\'e}chal}, \citenamefont {Gorceix}, \citenamefont {Rey}, \citenamefont {Vernac},\ and\ \citenamefont {Laburthe-Tolra}}]{Lepoutre2019ooe}%
  \BibitemOpen
  \bibfield  {author} {\bibinfo {author} {\bibfnamefont {S.}~\bibnamefont {Lepoutre}}, \bibinfo {author} {\bibfnamefont {J.}~\bibnamefont {Schachenmayer}}, \bibinfo {author} {\bibfnamefont {L.}~\bibnamefont {Gabardos}}, \bibinfo {author} {\bibfnamefont {B.}~\bibnamefont {Zhu}}, \bibinfo {author} {\bibfnamefont {B.}~\bibnamefont {Naylor}}, \bibinfo {author} {\bibfnamefont {E.}~\bibnamefont {Mar{\'e}chal}}, \bibinfo {author} {\bibfnamefont {O.}~\bibnamefont {Gorceix}}, \bibinfo {author} {\bibfnamefont {A.~M.}\ \bibnamefont {Rey}}, \bibinfo {author} {\bibfnamefont {L.}~\bibnamefont {Vernac}},\ and\ \bibinfo {author} {\bibfnamefont {B.}~\bibnamefont {Laburthe-Tolra}},\ }\bibfield  {title} {\bibinfo {title} {Out-of-equilibrium quantum magnetism and thermalization in a spin-3 many-body dipolar lattice system},\ }\href {https://doi.org/10.1038/s41467-019-09699-5} {\bibfield  {journal} {\bibinfo  {journal} {Nature Communications}\ }\textbf {\bibinfo {volume} {10}},\ \bibinfo {pages} {1714} (\bibinfo {year}
  {2019})}\BibitemShut {NoStop}%
\bibitem [{\citenamefont {Su}\ \emph {et~al.}(2023)\citenamefont {Su}, \citenamefont {Douglas}, \citenamefont {Szurek}, \citenamefont {Groth}, \citenamefont {Ozturk}, \citenamefont {Krahn}, \citenamefont {H{\'e}bert}, \citenamefont {Phelps}, \citenamefont {Ebadi}, \citenamefont {Dickerson}, \citenamefont {Ferlaino}, \citenamefont {Markovi{\'{c}}},\ and\ \citenamefont {Greiner}}]{Su2023dqs}%
  \BibitemOpen
  \bibfield  {author} {\bibinfo {author} {\bibfnamefont {L.}~\bibnamefont {Su}}, \bibinfo {author} {\bibfnamefont {A.}~\bibnamefont {Douglas}}, \bibinfo {author} {\bibfnamefont {M.}~\bibnamefont {Szurek}}, \bibinfo {author} {\bibfnamefont {R.}~\bibnamefont {Groth}}, \bibinfo {author} {\bibfnamefont {S.~F.}\ \bibnamefont {Ozturk}}, \bibinfo {author} {\bibfnamefont {A.}~\bibnamefont {Krahn}}, \bibinfo {author} {\bibfnamefont {A.~H.}\ \bibnamefont {H{\'e}bert}}, \bibinfo {author} {\bibfnamefont {G.~A.}\ \bibnamefont {Phelps}}, \bibinfo {author} {\bibfnamefont {S.}~\bibnamefont {Ebadi}}, \bibinfo {author} {\bibfnamefont {S.}~\bibnamefont {Dickerson}}, \bibinfo {author} {\bibfnamefont {F.}~\bibnamefont {Ferlaino}}, \bibinfo {author} {\bibfnamefont {O.}~\bibnamefont {Markovi{\'{c}}}},\ and\ \bibinfo {author} {\bibfnamefont {M.}~\bibnamefont {Greiner}},\ }\bibfield  {title} {\bibinfo {title} {Dipolar quantum solids emerging in a {H}ubbard quantum simulator},\ }\href {https://doi.org/10.1038/s41586-023-06614-3}
  {\bibfield  {journal} {\bibinfo  {journal} {Nature}\ }\textbf {\bibinfo {volume} {622}},\ \bibinfo {pages} {724} (\bibinfo {year} {2023})}\BibitemShut {NoStop}%
\bibitem [{\citenamefont {Mermin}\ and\ \citenamefont {Wagner}(1966)}]{Mermin1966}%
  \BibitemOpen
  \bibfield  {author} {\bibinfo {author} {\bibfnamefont {N.~D.}\ \bibnamefont {Mermin}}\ and\ \bibinfo {author} {\bibfnamefont {H.}~\bibnamefont {Wagner}},\ }\bibfield  {title} {\bibinfo {title} {Absence of ferromagnetism or antiferromagnetism in one- or two-dimensional isotropic {H}eisenberg models},\ }\href {https://doi.org/10.1103/PhysRevLett.17.1133} {\bibfield  {journal} {\bibinfo  {journal} {Phys. Rev. Lett.}\ }\textbf {\bibinfo {volume} {17}},\ \bibinfo {pages} {1133} (\bibinfo {year} {1966})}\BibitemShut {NoStop}%
\bibitem [{Note1()}]{Note1}%
  \BibitemOpen
  \bibinfo {note} {We expect our results to remain stable in the sector of total magnetization $S^z_\protect \text {tot} =\pm 1$. For higher/lower magnetization phase separated regimes, should occur. Here the system presents a region of zero total magnetization where the phase that we discuss persist, and one or more fully polarized Luttinger liquid regions~\cite {Montorsi2020}.}\BibitemShut {Stop}%
\bibitem [{\citenamefont {Giamarchi}(2003)}]{Giamarchi2003qpi}%
  \BibitemOpen
  \bibfield  {author} {\bibinfo {author} {\bibfnamefont {T.}~\bibnamefont {Giamarchi}},\ }\href@noop {} {\emph {\bibinfo {title} {Quantum physics in one dimension}}}\ (\bibinfo  {publisher} {Oxford Science Publications},\ \bibinfo {year} {2003})\BibitemShut {NoStop}%
\bibitem [{\citenamefont {Alexander O.~Gogolin}(1998)}]{Gogolin1998}%
  \BibitemOpen
  \bibfield  {author} {\bibinfo {author} {\bibfnamefont {A.~M.~T.}\ \bibnamefont {Alexander O.~Gogolin}, \bibfnamefont {Alexander A.~Nersesyan}},\ }\href@noop {} {\emph {\bibinfo {title} {Bosonization and Strongly Correlated Systems}}}\ (\bibinfo  {publisher} {Cambridge University Press},\ \bibinfo {year} {1998})\BibitemShut {NoStop}%
\bibitem [{\citenamefont {Nakamura}(2000)}]{Nakamura2000}%
  \BibitemOpen
  \bibfield  {author} {\bibinfo {author} {\bibfnamefont {M.}~\bibnamefont {Nakamura}},\ }\bibfield  {title} {\bibinfo {title} {Tricritical behavior in the extended hubbard chains},\ }\href {https://doi.org/10.1103/PhysRevB.61.16377} {\bibfield  {journal} {\bibinfo  {journal} {Phys. Rev. B}\ }\textbf {\bibinfo {volume} {61}},\ \bibinfo {pages} {16377} (\bibinfo {year} {2000})}\BibitemShut {NoStop}%
\bibitem [{\citenamefont {Barbiero}\ \emph {et~al.}(2013)\citenamefont {Barbiero}, \citenamefont {Montorsi},\ and\ \citenamefont {Roncaglia}}]{Barbiero2013hho}%
  \BibitemOpen
  \bibfield  {author} {\bibinfo {author} {\bibfnamefont {L.}~\bibnamefont {Barbiero}}, \bibinfo {author} {\bibfnamefont {A.}~\bibnamefont {Montorsi}},\ and\ \bibinfo {author} {\bibfnamefont {M.}~\bibnamefont {Roncaglia}},\ }\bibfield  {title} {\bibinfo {title} {How hidden orders generate gaps in one-dimensional fermionic systems},\ }\href {https://doi.org/10.1103/PhysRevB.88.035109} {\bibfield  {journal} {\bibinfo  {journal} {Phys. Rev. B}\ }\textbf {\bibinfo {volume} {88}},\ \bibinfo {pages} {035109} (\bibinfo {year} {2013})}\BibitemShut {NoStop}%
\bibitem [{\citenamefont {Baldelli}\ \emph {et~al.}(2024)\citenamefont {Baldelli}, \citenamefont {Montorsi}, \citenamefont {Julià-Farré}, \citenamefont {Lewenstein}, \citenamefont {Rizzi},\ and\ \citenamefont {Barbiero}}]{baldelli2024}%
  \BibitemOpen
  \bibfield  {author} {\bibinfo {author} {\bibfnamefont {N.}~\bibnamefont {Baldelli}}, \bibinfo {author} {\bibfnamefont {A.}~\bibnamefont {Montorsi}}, \bibinfo {author} {\bibfnamefont {S.}~\bibnamefont {Julià-Farré}}, \bibinfo {author} {\bibfnamefont {M.}~\bibnamefont {Lewenstein}}, \bibinfo {author} {\bibfnamefont {M.}~\bibnamefont {Rizzi}},\ and\ \bibinfo {author} {\bibfnamefont {L.}~\bibnamefont {Barbiero}},\ }\href {https://arxiv.org/abs/2407.04073} {\bibinfo {title} {Deconfined quantum critical points in fermionic systems with spin-charge separation}} (\bibinfo {year} {2024}),\ \Eprint {https://arxiv.org/abs/2407.04073} {arXiv:2407.04073 [cond-mat.str-el]} \BibitemShut {NoStop}%
\bibitem [{\citenamefont {Montorsi}\ and\ \citenamefont {Roncaglia}(2012)}]{Montorsi2012}%
  \BibitemOpen
  \bibfield  {author} {\bibinfo {author} {\bibfnamefont {A.}~\bibnamefont {Montorsi}}\ and\ \bibinfo {author} {\bibfnamefont {M.}~\bibnamefont {Roncaglia}},\ }\bibfield  {title} {\bibinfo {title} {Nonlocal order parameters for the {1D} {H}ubbard model},\ }\href {https://doi.org/10.1103/PhysRevLett.109.236404} {\bibfield  {journal} {\bibinfo  {journal} {Phys. Rev. Lett.}\ }\textbf {\bibinfo {volume} {109}},\ \bibinfo {pages} {236404} (\bibinfo {year} {2012})}\BibitemShut {NoStop}%
\bibitem [{\citenamefont {Fazzini}\ \emph {et~al.}(2019)\citenamefont {Fazzini}, \citenamefont {Barbiero},\ and\ \citenamefont {Montorsi}}]{Fazzini2019iif}%
  \BibitemOpen
  \bibfield  {author} {\bibinfo {author} {\bibfnamefont {S.}~\bibnamefont {Fazzini}}, \bibinfo {author} {\bibfnamefont {L.}~\bibnamefont {Barbiero}},\ and\ \bibinfo {author} {\bibfnamefont {A.}~\bibnamefont {Montorsi}},\ }\bibfield  {title} {\bibinfo {title} {Interaction-induced fractionalization and topological superconductivity in the polar molecules anisotropic $t\ensuremath{-}{J}$ model},\ }\href {https://doi.org/10.1103/PhysRevLett.122.106402} {\bibfield  {journal} {\bibinfo  {journal} {Phys. Rev. Lett.}\ }\textbf {\bibinfo {volume} {122}},\ \bibinfo {pages} {106402} (\bibinfo {year} {2019})}\BibitemShut {NoStop}%
\bibitem [{\citenamefont {Haldane}(1983)}]{Haldane1983}%
  \BibitemOpen
  \bibfield  {author} {\bibinfo {author} {\bibfnamefont {F.~D.~M.}\ \bibnamefont {Haldane}},\ }\bibfield  {title} {\bibinfo {title} {Nonlinear field theory of large-spin {H}eisenberg antiferromagnets: Semiclassically quantized solitons of the one-dimensional easy-axis {N}\'eel state},\ }\href {https://doi.org/10.1103/PhysRevLett.50.1153} {\bibfield  {journal} {\bibinfo  {journal} {Phys. Rev. Lett.}\ }\textbf {\bibinfo {volume} {50}},\ \bibinfo {pages} {1153} (\bibinfo {year} {1983})}\BibitemShut {NoStop}%
\bibitem [{\citenamefont {Pollmann}\ \emph {et~al.}(2012)\citenamefont {Pollmann}, \citenamefont {Berg}, \citenamefont {Turner},\ and\ \citenamefont {Oshikawa}}]{Pollmann2012}%
  \BibitemOpen
  \bibfield  {author} {\bibinfo {author} {\bibfnamefont {F.}~\bibnamefont {Pollmann}}, \bibinfo {author} {\bibfnamefont {E.}~\bibnamefont {Berg}}, \bibinfo {author} {\bibfnamefont {A.~M.}\ \bibnamefont {Turner}},\ and\ \bibinfo {author} {\bibfnamefont {M.}~\bibnamefont {Oshikawa}},\ }\bibfield  {title} {\bibinfo {title} {Symmetry protection of topological phases in one-dimensional quantum spin systems},\ }\href {https://doi.org/10.1103/PhysRevB.85.075125} {\bibfield  {journal} {\bibinfo  {journal} {Phys. Rev. B}\ }\textbf {\bibinfo {volume} {85}},\ \bibinfo {pages} {075125} (\bibinfo {year} {2012})}\BibitemShut {NoStop}%
\bibitem [{\citenamefont {Tang}\ and\ \citenamefont {Wen}(2012)}]{Tang2015}%
  \BibitemOpen
  \bibfield  {author} {\bibinfo {author} {\bibfnamefont {E.}~\bibnamefont {Tang}}\ and\ \bibinfo {author} {\bibfnamefont {X.-G.}\ \bibnamefont {Wen}},\ }\bibfield  {title} {\bibinfo {title} {Interacting one-dimensional fermionic symmetry-protected topological phases},\ }\href {https://doi.org/10.1103/PhysRevLett.109.096403} {\bibfield  {journal} {\bibinfo  {journal} {Phys. Rev. Lett.}\ }\textbf {\bibinfo {volume} {109}},\ \bibinfo {pages} {096403} (\bibinfo {year} {2012})}\BibitemShut {NoStop}%
\bibitem [{\citenamefont {Senthil}(2015)}]{Senthil2015}%
  \BibitemOpen
  \bibfield  {author} {\bibinfo {author} {\bibfnamefont {T.}~\bibnamefont {Senthil}},\ }\bibfield  {title} {\bibinfo {title} {Symmetry-protected topological phases of quantum matter},\ }\href {https://doi.org/https://doi.org/10.1146/annurev-conmatphys-031214-014740} {\bibfield  {journal} {\bibinfo  {journal} {Annual Review of Condensed Matter Physics}\ }\textbf {\bibinfo {volume} {6}},\ \bibinfo {pages} {299} (\bibinfo {year} {2015})}\BibitemShut {NoStop}%
\bibitem [{\citenamefont {Shapourian}\ \emph {et~al.}(2017)\citenamefont {Shapourian}, \citenamefont {Shiozaki},\ and\ \citenamefont {Ryu}}]{Shapourian2017}%
  \BibitemOpen
  \bibfield  {author} {\bibinfo {author} {\bibfnamefont {H.}~\bibnamefont {Shapourian}}, \bibinfo {author} {\bibfnamefont {K.}~\bibnamefont {Shiozaki}},\ and\ \bibinfo {author} {\bibfnamefont {S.}~\bibnamefont {Ryu}},\ }\bibfield  {title} {\bibinfo {title} {Many-body topological invariants for fermionic symmetry-protected topological phases},\ }\href {https://doi.org/10.1103/PhysRevLett.118.216402} {\bibfield  {journal} {\bibinfo  {journal} {Phys. Rev. Lett.}\ }\textbf {\bibinfo {volume} {118}},\ \bibinfo {pages} {216402} (\bibinfo {year} {2017})}\BibitemShut {NoStop}%
\bibitem [{\citenamefont {Montorsi}\ \emph {et~al.}(2017)\citenamefont {Montorsi}, \citenamefont {Dolcini}, \citenamefont {Iotti},\ and\ \citenamefont {Rossi}}]{Montorsi2017}%
  \BibitemOpen
  \bibfield  {author} {\bibinfo {author} {\bibfnamefont {A.}~\bibnamefont {Montorsi}}, \bibinfo {author} {\bibfnamefont {F.}~\bibnamefont {Dolcini}}, \bibinfo {author} {\bibfnamefont {R.~C.}\ \bibnamefont {Iotti}},\ and\ \bibinfo {author} {\bibfnamefont {F.}~\bibnamefont {Rossi}},\ }\bibfield  {title} {\bibinfo {title} {Symmetry-protected topological phases of one-dimensional interacting fermions with spin-charge separation},\ }\href {https://doi.org/10.1103/PhysRevB.95.245108} {\bibfield  {journal} {\bibinfo  {journal} {Phys. Rev. B}\ }\textbf {\bibinfo {volume} {95}},\ \bibinfo {pages} {245108} (\bibinfo {year} {2017})}\BibitemShut {NoStop}%
\bibitem [{\citenamefont {Jackiw}\ and\ \citenamefont {Rebbi}(1976)}]{Jackiw1976swf}%
  \BibitemOpen
  \bibfield  {author} {\bibinfo {author} {\bibfnamefont {R.}~\bibnamefont {Jackiw}}\ and\ \bibinfo {author} {\bibfnamefont {C.}~\bibnamefont {Rebbi}},\ }\bibfield  {title} {\bibinfo {title} {Solitons with fermion number \textonehalf{}},\ }\href {https://doi.org/10.1103/PhysRevD.13.3398} {\bibfield  {journal} {\bibinfo  {journal} {Phys. Rev. D}\ }\textbf {\bibinfo {volume} {13}},\ \bibinfo {pages} {3398} (\bibinfo {year} {1976})}\BibitemShut {NoStop}%
\bibitem [{\citenamefont {Su}\ \emph {et~al.}(1979)\citenamefont {Su}, \citenamefont {Schrieffer},\ and\ \citenamefont {Heeger}}]{Su1979sip}%
  \BibitemOpen
  \bibfield  {author} {\bibinfo {author} {\bibfnamefont {W.~P.}\ \bibnamefont {Su}}, \bibinfo {author} {\bibfnamefont {J.~R.}\ \bibnamefont {Schrieffer}},\ and\ \bibinfo {author} {\bibfnamefont {A.~J.}\ \bibnamefont {Heeger}},\ }\bibfield  {title} {\bibinfo {title} {Solitons in polyacetylene},\ }\href {https://doi.org/10.1103/PhysRevLett.42.1698} {\bibfield  {journal} {\bibinfo  {journal} {Phys. Rev. Lett.}\ }\textbf {\bibinfo {volume} {42}},\ \bibinfo {pages} {1698} (\bibinfo {year} {1979})}\BibitemShut {NoStop}%
\bibitem [{\citenamefont {Fabrizio}\ and\ \citenamefont {Gogolin}(1995)}]{Fabrizio1995iod}%
  \BibitemOpen
  \bibfield  {author} {\bibinfo {author} {\bibfnamefont {M.}~\bibnamefont {Fabrizio}}\ and\ \bibinfo {author} {\bibfnamefont {A.~O.}\ \bibnamefont {Gogolin}},\ }\bibfield  {title} {\bibinfo {title} {Interacting one-dimensional electron gas with open boundaries},\ }\href {https://doi.org/10.1103/PhysRevB.51.17827} {\bibfield  {journal} {\bibinfo  {journal} {Phys. Rev. B}\ }\textbf {\bibinfo {volume} {51}},\ \bibinfo {pages} {17827} (\bibinfo {year} {1995})}\BibitemShut {NoStop}%
\bibitem [{\citenamefont {White}(1992)}]{White1992}%
  \BibitemOpen
  \bibfield  {author} {\bibinfo {author} {\bibfnamefont {S.~R.}\ \bibnamefont {White}},\ }\bibfield  {title} {\bibinfo {title} {Density matrix formulation for quantum renormalization groups},\ }\href {https://doi.org/10.1103/PhysRevLett.69.2863} {\bibfield  {journal} {\bibinfo  {journal} {Phys. Rev. Lett.}\ }\textbf {\bibinfo {volume} {69}},\ \bibinfo {pages} {2863} (\bibinfo {year} {1992})}\BibitemShut {NoStop}%
\bibitem [{\citenamefont {Schollw{\"o}ck}(2011)}]{SCHOLLWOCK201196}%
  \BibitemOpen
  \bibfield  {author} {\bibinfo {author} {\bibfnamefont {U.}~\bibnamefont {Schollw{\"o}ck}},\ }\bibfield  {title} {\bibinfo {title} {The density-matrix renormalization group in the age of matrix product states},\ }\href {https://doi.org/https://doi.org/10.1016/j.aop.2010.09.012} {\bibfield  {journal} {\bibinfo  {journal} {Annals of Physics}\ }\textbf {\bibinfo {volume} {326}},\ \bibinfo {pages} {96} (\bibinfo {year} {2011})}\BibitemShut {NoStop}%
\bibitem [{\citenamefont {Luther}\ and\ \citenamefont {Emery}(1974)}]{Luther1974}%
  \BibitemOpen
  \bibfield  {author} {\bibinfo {author} {\bibfnamefont {A.}~\bibnamefont {Luther}}\ and\ \bibinfo {author} {\bibfnamefont {V.~J.}\ \bibnamefont {Emery}},\ }\bibfield  {title} {\bibinfo {title} {Backward scattering in the one-dimensional electron gas},\ }\href {https://doi.org/10.1103/PhysRevLett.33.589} {\bibfield  {journal} {\bibinfo  {journal} {Phys. Rev. Lett.}\ }\textbf {\bibinfo {volume} {33}},\ \bibinfo {pages} {589} (\bibinfo {year} {1974})}\BibitemShut {NoStop}%
\bibitem [{\citenamefont {Li}\ and\ \citenamefont {Haldane}(2008)}]{Li2008}%
  \BibitemOpen
  \bibfield  {author} {\bibinfo {author} {\bibfnamefont {H.}~\bibnamefont {Li}}\ and\ \bibinfo {author} {\bibfnamefont {F.~D.~M.}\ \bibnamefont {Haldane}},\ }\bibfield  {title} {\bibinfo {title} {Entanglement spectrum as a generalization of entanglement entropy: Identification of topological order in non-abelian fractional quantum {H}all effect states},\ }\href {https://doi.org/10.1103/PhysRevLett.101.010504} {\bibfield  {journal} {\bibinfo  {journal} {Phys. Rev. Lett.}\ }\textbf {\bibinfo {volume} {101}},\ \bibinfo {pages} {010504} (\bibinfo {year} {2008})}\BibitemShut {NoStop}%
\bibitem [{\citenamefont {Pollmann}\ \emph {et~al.}(2010)\citenamefont {Pollmann}, \citenamefont {Turner}, \citenamefont {Berg},\ and\ \citenamefont {Oshikawa}}]{Pollmann2010}%
  \BibitemOpen
  \bibfield  {author} {\bibinfo {author} {\bibfnamefont {F.}~\bibnamefont {Pollmann}}, \bibinfo {author} {\bibfnamefont {A.~M.}\ \bibnamefont {Turner}}, \bibinfo {author} {\bibfnamefont {E.}~\bibnamefont {Berg}},\ and\ \bibinfo {author} {\bibfnamefont {M.}~\bibnamefont {Oshikawa}},\ }\bibfield  {title} {\bibinfo {title} {Entanglement spectrum of a topological phase in one dimension},\ }\href {https://doi.org/10.1103/PhysRevB.81.064439} {\bibfield  {journal} {\bibinfo  {journal} {Phys. Rev. B}\ }\textbf {\bibinfo {volume} {81}},\ \bibinfo {pages} {064439} (\bibinfo {year} {2010})}\BibitemShut {NoStop}%
\bibitem [{\citenamefont {Turner}\ \emph {et~al.}(2011)\citenamefont {Turner}, \citenamefont {Pollmann},\ and\ \citenamefont {Berg}}]{Turner2011}%
  \BibitemOpen
  \bibfield  {author} {\bibinfo {author} {\bibfnamefont {A.~M.}\ \bibnamefont {Turner}}, \bibinfo {author} {\bibfnamefont {F.}~\bibnamefont {Pollmann}},\ and\ \bibinfo {author} {\bibfnamefont {E.}~\bibnamefont {Berg}},\ }\bibfield  {title} {\bibinfo {title} {Topological phases of one-dimensional fermions: An entanglement point of view},\ }\href {https://doi.org/10.1103/PhysRevB.83.075102} {\bibfield  {journal} {\bibinfo  {journal} {Phys. Rev. B}\ }\textbf {\bibinfo {volume} {83}},\ \bibinfo {pages} {075102} (\bibinfo {year} {2011})}\BibitemShut {NoStop}%
\bibitem [{\citenamefont {de~L{\'e}s{\'e}leuc}\ \emph {et~al.}(2019)\citenamefont {de~L{\'e}s{\'e}leuc}, \citenamefont {Lienhard}, \citenamefont {Scholl}, \citenamefont {Barredo}, \citenamefont {Weber}, \citenamefont {Lang}, \citenamefont {B{\"u}chler}, \citenamefont {Lahaye},\ and\ \citenamefont {Browaeys}}]{Leseleuc2019}%
  \BibitemOpen
  \bibfield  {author} {\bibinfo {author} {\bibfnamefont {S.}~\bibnamefont {de~L{\'e}s{\'e}leuc}}, \bibinfo {author} {\bibfnamefont {V.}~\bibnamefont {Lienhard}}, \bibinfo {author} {\bibfnamefont {P.}~\bibnamefont {Scholl}}, \bibinfo {author} {\bibfnamefont {D.}~\bibnamefont {Barredo}}, \bibinfo {author} {\bibfnamefont {S.}~\bibnamefont {Weber}}, \bibinfo {author} {\bibfnamefont {N.}~\bibnamefont {Lang}}, \bibinfo {author} {\bibfnamefont {H.~P.}\ \bibnamefont {B{\"u}chler}}, \bibinfo {author} {\bibfnamefont {T.}~\bibnamefont {Lahaye}},\ and\ \bibinfo {author} {\bibfnamefont {A.}~\bibnamefont {Browaeys}},\ }\bibfield  {title} {\bibinfo {title} {Observation of a symmetry-protected topological phase of interacting bosons with {R}ydberg atoms},\ }\href {https://doi.org/10.1126/science.aav9105} {\bibfield  {journal} {\bibinfo  {journal} {Science}\ }\textbf {\bibinfo {volume} {365}},\ \bibinfo {pages} {775} (\bibinfo {year} {2019})}\BibitemShut {NoStop}%
\bibitem [{\citenamefont {Su}\ \emph {et~al.}(2025{\natexlab{a}})\citenamefont {Su}, \citenamefont {Sahay}, \citenamefont {Szurek}, \citenamefont {Douglas}, \citenamefont {Markovic}, \citenamefont {Dag}, \citenamefont {Verresen},\ and\ \citenamefont {Greiner}}]{su2025topological}%
  \BibitemOpen
  \bibfield  {author} {\bibinfo {author} {\bibfnamefont {L.}~\bibnamefont {Su}}, \bibinfo {author} {\bibfnamefont {R.}~\bibnamefont {Sahay}}, \bibinfo {author} {\bibfnamefont {M.}~\bibnamefont {Szurek}}, \bibinfo {author} {\bibfnamefont {A.}~\bibnamefont {Douglas}}, \bibinfo {author} {\bibfnamefont {O.}~\bibnamefont {Markovic}}, \bibinfo {author} {\bibfnamefont {C.~B.}\ \bibnamefont {Dag}}, \bibinfo {author} {\bibfnamefont {R.}~\bibnamefont {Verresen}},\ and\ \bibinfo {author} {\bibfnamefont {M.}~\bibnamefont {Greiner}},\ }\bibfield  {title} {\bibinfo {title} {Topological phases, criticality, and mixed state order in a {H}ubbard quantum simulator},\ }\href {https://arxiv.org/abs/2505.17009} {\bibfield  {journal} {\bibinfo  {journal} {arXiv preprint arXiv:2505.17009}\ } (\bibinfo {year} {2025}{\natexlab{a}})}\BibitemShut {NoStop}%
\bibitem [{\citenamefont {Montorsi}\ \emph {et~al.}(2020)\citenamefont {Montorsi}, \citenamefont {Fazzini},\ and\ \citenamefont {Barbiero}}]{Montorsi2020}%
  \BibitemOpen
  \bibfield  {author} {\bibinfo {author} {\bibfnamefont {A.}~\bibnamefont {Montorsi}}, \bibinfo {author} {\bibfnamefont {S.}~\bibnamefont {Fazzini}},\ and\ \bibinfo {author} {\bibfnamefont {L.}~\bibnamefont {Barbiero}},\ }\bibfield  {title} {\bibinfo {title} {Homogeneous and domain-wall topological {H}aldane conductors with dressed {R}ydberg atoms},\ }\href {https://doi.org/10.1103/PhysRevA.101.043618} {\bibfield  {journal} {\bibinfo  {journal} {Phys. Rev. A}\ }\textbf {\bibinfo {volume} {101}},\ \bibinfo {pages} {043618} (\bibinfo {year} {2020})}\BibitemShut {NoStop}%
\bibitem [{\citenamefont {Guardado-Sanchez}\ \emph {et~al.}(2020)\citenamefont {Guardado-Sanchez}, \citenamefont {Morningstar}, \citenamefont {Spar}, \citenamefont {Brown}, \citenamefont {Huse},\ and\ \citenamefont {Bakr}}]{Guardado2020sah}%
  \BibitemOpen
  \bibfield  {author} {\bibinfo {author} {\bibfnamefont {E.}~\bibnamefont {Guardado-Sanchez}}, \bibinfo {author} {\bibfnamefont {A.}~\bibnamefont {Morningstar}}, \bibinfo {author} {\bibfnamefont {B.~M.}\ \bibnamefont {Spar}}, \bibinfo {author} {\bibfnamefont {P.~T.}\ \bibnamefont {Brown}}, \bibinfo {author} {\bibfnamefont {D.~A.}\ \bibnamefont {Huse}},\ and\ \bibinfo {author} {\bibfnamefont {W.~S.}\ \bibnamefont {Bakr}},\ }\bibfield  {title} {\bibinfo {title} {Subdiffusion and heat transport in a tilted two-dimensional {F}ermi-{H}ubbard system},\ }\href {https://doi.org/10.1103/PhysRevX.10.011042} {\bibfield  {journal} {\bibinfo  {journal} {Phys. Rev. X}\ }\textbf {\bibinfo {volume} {10}},\ \bibinfo {pages} {011042} (\bibinfo {year} {2020})}\BibitemShut {NoStop}%
\bibitem [{\citenamefont {Weitenberg}\ \emph {et~al.}(2011)\citenamefont {Weitenberg}, \citenamefont {Endres}, \citenamefont {Sherson}, \citenamefont {Cheneau}, \citenamefont {Schau{\ss}}, \citenamefont {Fukuhara}, \citenamefont {Bloch},\ and\ \citenamefont {Kuhr}}]{Weitenberg2011ssa}%
  \BibitemOpen
  \bibfield  {author} {\bibinfo {author} {\bibfnamefont {C.}~\bibnamefont {Weitenberg}}, \bibinfo {author} {\bibfnamefont {M.}~\bibnamefont {Endres}}, \bibinfo {author} {\bibfnamefont {J.~F.}\ \bibnamefont {Sherson}}, \bibinfo {author} {\bibfnamefont {M.}~\bibnamefont {Cheneau}}, \bibinfo {author} {\bibfnamefont {P.}~\bibnamefont {Schau{\ss}}}, \bibinfo {author} {\bibfnamefont {T.}~\bibnamefont {Fukuhara}}, \bibinfo {author} {\bibfnamefont {I.}~\bibnamefont {Bloch}},\ and\ \bibinfo {author} {\bibfnamefont {S.}~\bibnamefont {Kuhr}},\ }\bibfield  {title} {\bibinfo {title} {Single-spin addressing in an atomic {M}ott insulator},\ }\href {https://doi.org/10.1038/nature09827} {\bibfield  {journal} {\bibinfo  {journal} {Nature}\ }\textbf {\bibinfo {volume} {471}},\ \bibinfo {pages} {319} (\bibinfo {year} {2011})}\BibitemShut {NoStop}%
\bibitem [{\citenamefont {Sohmen}\ \emph {et~al.}(2023)\citenamefont {Sohmen}, \citenamefont {Mark}, \citenamefont {Greiner},\ and\ \citenamefont {Ferlaino}}]{Sohmen2023asi}%
  \BibitemOpen
  \bibfield  {author} {\bibinfo {author} {\bibfnamefont {M.}~\bibnamefont {Sohmen}}, \bibinfo {author} {\bibfnamefont {M.~J.}\ \bibnamefont {Mark}}, \bibinfo {author} {\bibfnamefont {M.}~\bibnamefont {Greiner}},\ and\ \bibinfo {author} {\bibfnamefont {F.}~\bibnamefont {Ferlaino}},\ }\bibfield  {title} {\bibinfo {title} {{A ship-in-a-bottle quantum gas microscope setup for magnetic mixtures}},\ }\href {https://doi.org/10.21468/SciPostPhys.15.5.182} {\bibfield  {journal} {\bibinfo  {journal} {SciPost Phys.}\ }\textbf {\bibinfo {volume} {15}},\ \bibinfo {pages} {182} (\bibinfo {year} {2023})}\BibitemShut {NoStop}%
\bibitem [{\citenamefont {Picard}\ \emph {et~al.}(2019)\citenamefont {Picard}, \citenamefont {Mark}, \citenamefont {Ferlaino},\ and\ \citenamefont {van Bijnen}}]{Picard2019dla}%
  \BibitemOpen
  \bibfield  {author} {\bibinfo {author} {\bibfnamefont {L.~R.~B.}\ \bibnamefont {Picard}}, \bibinfo {author} {\bibfnamefont {M.~J.}\ \bibnamefont {Mark}}, \bibinfo {author} {\bibfnamefont {F.}~\bibnamefont {Ferlaino}},\ and\ \bibinfo {author} {\bibfnamefont {R.}~\bibnamefont {van Bijnen}},\ }\bibfield  {title} {\bibinfo {title} {Deep learning-assisted classification of site-resolved quantum gas microscope images},\ }\href {https://doi.org/10.1088/1361-6501/ab44d8} {\bibfield  {journal} {\bibinfo  {journal} {Measurement Science and Technology}\ }\textbf {\bibinfo {volume} {31}},\ \bibinfo {pages} {025201} (\bibinfo {year} {2019})}\BibitemShut {NoStop}%
\bibitem [{\citenamefont {Impertro}\ \emph {et~al.}(2023)\citenamefont {Impertro}, \citenamefont {Wienand}, \citenamefont {H{\"a}fele}, \citenamefont {von Raven}, \citenamefont {Hubele}, \citenamefont {Klostermann}, \citenamefont {Cabrera}, \citenamefont {Bloch},\ and\ \citenamefont {Aidelsburger}}]{Impertro2023aud}%
  \BibitemOpen
  \bibfield  {author} {\bibinfo {author} {\bibfnamefont {A.}~\bibnamefont {Impertro}}, \bibinfo {author} {\bibfnamefont {J.~F.}\ \bibnamefont {Wienand}}, \bibinfo {author} {\bibfnamefont {S.}~\bibnamefont {H{\"a}fele}}, \bibinfo {author} {\bibfnamefont {H.}~\bibnamefont {von Raven}}, \bibinfo {author} {\bibfnamefont {S.}~\bibnamefont {Hubele}}, \bibinfo {author} {\bibfnamefont {T.}~\bibnamefont {Klostermann}}, \bibinfo {author} {\bibfnamefont {C.~R.}\ \bibnamefont {Cabrera}}, \bibinfo {author} {\bibfnamefont {I.}~\bibnamefont {Bloch}},\ and\ \bibinfo {author} {\bibfnamefont {M.}~\bibnamefont {Aidelsburger}},\ }\bibfield  {title} {\bibinfo {title} {An unsupervised deep learning algorithm for single-site reconstruction in quantum gas microscopes},\ }\href {https://doi.org/10.1038/s42005-023-01287-w} {\bibfield  {journal} {\bibinfo  {journal} {Communications Physics}\ }\textbf {\bibinfo {volume} {6}},\ \bibinfo {pages} {166} (\bibinfo {year} {2023})}\BibitemShut {NoStop}%
\bibitem [{\citenamefont {Asteria}\ \emph {et~al.}(2021)\citenamefont {Asteria}, \citenamefont {Zahn}, \citenamefont {Kosch}, \citenamefont {Sengstock},\ and\ \citenamefont {Weitenberg}}]{Asteria2021qgm}%
  \BibitemOpen
  \bibfield  {author} {\bibinfo {author} {\bibfnamefont {L.}~\bibnamefont {Asteria}}, \bibinfo {author} {\bibfnamefont {H.~P.}\ \bibnamefont {Zahn}}, \bibinfo {author} {\bibfnamefont {M.~N.}\ \bibnamefont {Kosch}}, \bibinfo {author} {\bibfnamefont {K.}~\bibnamefont {Sengstock}},\ and\ \bibinfo {author} {\bibfnamefont {C.}~\bibnamefont {Weitenberg}},\ }\bibfield  {title} {\bibinfo {title} {Quantum gas magnifier for sub-lattice-resolved imaging of {3D} quantum systems},\ }\href {https://doi.org/10.1038/s41586-021-04011-2} {\bibfield  {journal} {\bibinfo  {journal} {Nature}\ }\textbf {\bibinfo {volume} {599}},\ \bibinfo {pages} {571} (\bibinfo {year} {2021})}\BibitemShut {NoStop}%
\bibitem [{\citenamefont {Su}\ \emph {et~al.}(2025{\natexlab{b}})\citenamefont {Su}, \citenamefont {Douglas}, \citenamefont {Szurek}, \citenamefont {H{\'e}bert}, \citenamefont {Krahn}, \citenamefont {Groth}, \citenamefont {Phelps}, \citenamefont {Markovi{\'{c}}},\ and\ \citenamefont {Greiner}}]{Su2025hrf}%
  \BibitemOpen
  \bibfield  {author} {\bibinfo {author} {\bibfnamefont {L.}~\bibnamefont {Su}}, \bibinfo {author} {\bibfnamefont {A.}~\bibnamefont {Douglas}}, \bibinfo {author} {\bibfnamefont {M.}~\bibnamefont {Szurek}}, \bibinfo {author} {\bibfnamefont {A.~H.}\ \bibnamefont {H{\'e}bert}}, \bibinfo {author} {\bibfnamefont {A.}~\bibnamefont {Krahn}}, \bibinfo {author} {\bibfnamefont {R.}~\bibnamefont {Groth}}, \bibinfo {author} {\bibfnamefont {G.~A.}\ \bibnamefont {Phelps}}, \bibinfo {author} {\bibfnamefont {O.}~\bibnamefont {Markovi{\'{c}}}},\ and\ \bibinfo {author} {\bibfnamefont {M.}~\bibnamefont {Greiner}},\ }\bibfield  {title} {\bibinfo {title} {Fast single atom imaging for optical lattice arrays},\ }\href {https://doi.org/10.1038/s41467-025-56305-y} {\bibfield  {journal} {\bibinfo  {journal} {Nature Communications}\ }\textbf {\bibinfo {volume} {16}},\ \bibinfo {pages} {1017} (\bibinfo {year} {2025}{\natexlab{b}})}\BibitemShut {NoStop}%
\bibitem [{\citenamefont {Boll}\ \emph {et~al.}(2016)\citenamefont {Boll}, \citenamefont {Hilker}, \citenamefont {Salomon}, \citenamefont {Omran}, \citenamefont {Nespolo}, \citenamefont {Pollet}, \citenamefont {Bloch},\ and\ \citenamefont {Gross}}]{Boll2016sad}%
  \BibitemOpen
  \bibfield  {author} {\bibinfo {author} {\bibfnamefont {M.}~\bibnamefont {Boll}}, \bibinfo {author} {\bibfnamefont {T.~A.}\ \bibnamefont {Hilker}}, \bibinfo {author} {\bibfnamefont {G.}~\bibnamefont {Salomon}}, \bibinfo {author} {\bibfnamefont {A.}~\bibnamefont {Omran}}, \bibinfo {author} {\bibfnamefont {J.}~\bibnamefont {Nespolo}}, \bibinfo {author} {\bibfnamefont {L.}~\bibnamefont {Pollet}}, \bibinfo {author} {\bibfnamefont {I.}~\bibnamefont {Bloch}},\ and\ \bibinfo {author} {\bibfnamefont {C.}~\bibnamefont {Gross}},\ }\bibfield  {title} {\bibinfo {title} {Spin- and density-resolved microscopy of antiferromagnetic correlations in {F}ermi-{H}ubbard chains},\ }\href {https://doi.org/10.1126/science.aag1635} {\bibfield  {journal} {\bibinfo  {journal} {Science}\ }\textbf {\bibinfo {volume} {353}},\ \bibinfo {pages} {1257} (\bibinfo {year} {2016})}\BibitemShut {NoStop}%
\bibitem [{\citenamefont {Impertro}\ \emph {et~al.}(2024)\citenamefont {Impertro}, \citenamefont {Karch}, \citenamefont {Wienand}, \citenamefont {Huh}, \citenamefont {Schweizer}, \citenamefont {Bloch},\ and\ \citenamefont {Aidelsburger}}]{impertro2024lra}%
  \BibitemOpen
  \bibfield  {author} {\bibinfo {author} {\bibfnamefont {A.}~\bibnamefont {Impertro}}, \bibinfo {author} {\bibfnamefont {S.}~\bibnamefont {Karch}}, \bibinfo {author} {\bibfnamefont {J.~F.}\ \bibnamefont {Wienand}}, \bibinfo {author} {\bibfnamefont {S.}~\bibnamefont {Huh}}, \bibinfo {author} {\bibfnamefont {C.}~\bibnamefont {Schweizer}}, \bibinfo {author} {\bibfnamefont {I.}~\bibnamefont {Bloch}},\ and\ \bibinfo {author} {\bibfnamefont {M.}~\bibnamefont {Aidelsburger}},\ }\bibfield  {title} {\bibinfo {title} {Local readout and control of current and kinetic energy operators in optical lattices},\ }\href {https://doi.org/10.1103/PhysRevLett.133.063401} {\bibfield  {journal} {\bibinfo  {journal} {Phys. Rev. Lett.}\ }\textbf {\bibinfo {volume} {133}},\ \bibinfo {pages} {063401} (\bibinfo {year} {2024})}\BibitemShut {NoStop}%
\bibitem [{\citenamefont {Mark}\ \emph {et~al.}(2025)\citenamefont {Mark}, \citenamefont {Hu}, \citenamefont {Kwan}, \citenamefont {Kokail}, \citenamefont {Choi},\ and\ \citenamefont {Yelin}}]{mark2025emd}%
  \BibitemOpen
  \bibfield  {author} {\bibinfo {author} {\bibfnamefont {D.~K.}\ \bibnamefont {Mark}}, \bibinfo {author} {\bibfnamefont {H.-Y.}\ \bibnamefont {Hu}}, \bibinfo {author} {\bibfnamefont {J.}~\bibnamefont {Kwan}}, \bibinfo {author} {\bibfnamefont {C.}~\bibnamefont {Kokail}}, \bibinfo {author} {\bibfnamefont {S.}~\bibnamefont {Choi}},\ and\ \bibinfo {author} {\bibfnamefont {S.~F.}\ \bibnamefont {Yelin}},\ }\bibfield  {title} {\bibinfo {title} {Efficiently measuring $d$-wave pairing and beyond in quantum gas microscopes},\ }\href {https://doi.org/10.1103/dqyf-kl8x} {\bibfield  {journal} {\bibinfo  {journal} {Phys. Rev. Lett.}\ }\textbf {\bibinfo {volume} {135}},\ \bibinfo {pages} {123402} (\bibinfo {year} {2025})}\BibitemShut {NoStop}%
\bibitem [{\citenamefont {Strinati}\ \emph {et~al.}(2018)\citenamefont {Strinati}, \citenamefont {Gerbier},\ and\ \citenamefont {Mazza}}]{Strinati2018sgs}%
  \BibitemOpen
  \bibfield  {author} {\bibinfo {author} {\bibfnamefont {M.~C.}\ \bibnamefont {Strinati}}, \bibinfo {author} {\bibfnamefont {F.}~\bibnamefont {Gerbier}},\ and\ \bibinfo {author} {\bibfnamefont {L.}~\bibnamefont {Mazza}},\ }\bibfield  {title} {\bibinfo {title} {Spin-gap spectroscopy in a bosonic flux ladder},\ }\href {https://doi.org/10.1088/1367-2630/aa9ca2} {\bibfield  {journal} {\bibinfo  {journal} {New Journal of Physics}\ }\textbf {\bibinfo {volume} {20}},\ \bibinfo {pages} {015004} (\bibinfo {year} {2018})}\BibitemShut {NoStop}%
\bibitem [{\citenamefont {Pichler}\ \emph {et~al.}(2016)\citenamefont {Pichler}, \citenamefont {Zhu}, \citenamefont {Seif}, \citenamefont {Zoller},\ and\ \citenamefont {Hafezi}}]{pichler2016mpf}%
  \BibitemOpen
  \bibfield  {author} {\bibinfo {author} {\bibfnamefont {H.}~\bibnamefont {Pichler}}, \bibinfo {author} {\bibfnamefont {G.}~\bibnamefont {Zhu}}, \bibinfo {author} {\bibfnamefont {A.}~\bibnamefont {Seif}}, \bibinfo {author} {\bibfnamefont {P.}~\bibnamefont {Zoller}},\ and\ \bibinfo {author} {\bibfnamefont {M.}~\bibnamefont {Hafezi}},\ }\bibfield  {title} {\bibinfo {title} {Measurement protocol for the entanglement spectrum of cold atoms},\ }\href {https://doi.org/10.1103/PhysRevX.6.041033} {\bibfield  {journal} {\bibinfo  {journal} {Phys. Rev. X}\ }\textbf {\bibinfo {volume} {6}},\ \bibinfo {pages} {041033} (\bibinfo {year} {2016})}\BibitemShut {NoStop}%
\bibitem [{\citenamefont {Zheng}\ \emph {et~al.}(2017)\citenamefont {Zheng}, \citenamefont {Chung}, \citenamefont {Corboz}, \citenamefont {Ehlers}, \citenamefont {Qin}, \citenamefont {Noack}, \citenamefont {Shi}, \citenamefont {White}, \citenamefont {Zhang},\ and\ \citenamefont {Chan}}]{Bo-Xiao2017}%
  \BibitemOpen
  \bibfield  {author} {\bibinfo {author} {\bibfnamefont {B.-X.}\ \bibnamefont {Zheng}}, \bibinfo {author} {\bibfnamefont {C.-M.}\ \bibnamefont {Chung}}, \bibinfo {author} {\bibfnamefont {P.}~\bibnamefont {Corboz}}, \bibinfo {author} {\bibfnamefont {G.}~\bibnamefont {Ehlers}}, \bibinfo {author} {\bibfnamefont {M.-P.}\ \bibnamefont {Qin}}, \bibinfo {author} {\bibfnamefont {R.~M.}\ \bibnamefont {Noack}}, \bibinfo {author} {\bibfnamefont {H.}~\bibnamefont {Shi}}, \bibinfo {author} {\bibfnamefont {S.~R.}\ \bibnamefont {White}}, \bibinfo {author} {\bibfnamefont {S.}~\bibnamefont {Zhang}},\ and\ \bibinfo {author} {\bibfnamefont {G.~K.-L.}\ \bibnamefont {Chan}},\ }\bibfield  {title} {\bibinfo {title} {Stripe order in the underdoped region of the two-dimensional {H}ubbard model},\ }\href {https://doi.org/10.1126/science.aam7127} {\bibfield  {journal} {\bibinfo  {journal} {Science}\ }\textbf {\bibinfo {volume} {358}},\ \bibinfo {pages} {1155} (\bibinfo {year} {2017})}\BibitemShut {NoStop}%
\bibitem [{\citenamefont {Wietek}\ \emph {et~al.}(2021)\citenamefont {Wietek}, \citenamefont {He}, \citenamefont {White}, \citenamefont {Georges},\ and\ \citenamefont {Stoudenmire}}]{Wietek2021}%
  \BibitemOpen
  \bibfield  {author} {\bibinfo {author} {\bibfnamefont {A.}~\bibnamefont {Wietek}}, \bibinfo {author} {\bibfnamefont {Y.-Y.}\ \bibnamefont {He}}, \bibinfo {author} {\bibfnamefont {S.~R.}\ \bibnamefont {White}}, \bibinfo {author} {\bibfnamefont {A.}~\bibnamefont {Georges}},\ and\ \bibinfo {author} {\bibfnamefont {E.~M.}\ \bibnamefont {Stoudenmire}},\ }\bibfield  {title} {\bibinfo {title} {Stripes, antiferromagnetism, and the pseudogap in the doped {H}ubbard model at finite temperature},\ }\href {https://doi.org/10.1103/PhysRevX.11.031007} {\bibfield  {journal} {\bibinfo  {journal} {Phys. Rev. X}\ }\textbf {\bibinfo {volume} {11}},\ \bibinfo {pages} {031007} (\bibinfo {year} {2021})}\BibitemShut {NoStop}%
\bibitem [{\citenamefont {Xu}\ \emph {et~al.}(2025)\citenamefont {Xu}, \citenamefont {Kendrick}, \citenamefont {Kale}, \citenamefont {Gang}, \citenamefont {Feng}, \citenamefont {Zhang}, \citenamefont {Young}, \citenamefont {Lebrat},\ and\ \citenamefont {Greiner}}]{Xu2025}%
  \BibitemOpen
  \bibfield  {author} {\bibinfo {author} {\bibfnamefont {M.}~\bibnamefont {Xu}}, \bibinfo {author} {\bibfnamefont {L.~H.}\ \bibnamefont {Kendrick}}, \bibinfo {author} {\bibfnamefont {A.}~\bibnamefont {Kale}}, \bibinfo {author} {\bibfnamefont {Y.}~\bibnamefont {Gang}}, \bibinfo {author} {\bibfnamefont {C.}~\bibnamefont {Feng}}, \bibinfo {author} {\bibfnamefont {S.}~\bibnamefont {Zhang}}, \bibinfo {author} {\bibfnamefont {A.~W.}\ \bibnamefont {Young}}, \bibinfo {author} {\bibfnamefont {M.}~\bibnamefont {Lebrat}},\ and\ \bibinfo {author} {\bibfnamefont {M.}~\bibnamefont {Greiner}},\ }\bibfield  {title} {\bibinfo {title} {A neutral-atom {H}ubbard quantum simulator in the cryogenic regime},\ }\href {https://doi.org/10.1038/s41586-025-09112-w} {\bibfield  {journal} {\bibinfo  {journal} {Nature}\ }\textbf {\bibinfo {volume} {642}},\ \bibinfo {pages} {909} (\bibinfo {year} {2025})}\BibitemShut {NoStop}%
\bibitem [{\citenamefont {Kitaev}(2001)}]{Kitaev_2001}%
  \BibitemOpen
  \bibfield  {author} {\bibinfo {author} {\bibfnamefont {A.~Y.}\ \bibnamefont {Kitaev}},\ }\bibfield  {title} {\bibinfo {title} {Unpaired {M}ajorana fermions in quantum wires},\ }\href {https://doi.org/10.1070/1063-7869/44/10S/S29} {\bibfield  {journal} {\bibinfo  {journal} {Physics-Uspekhi}\ }\textbf {\bibinfo {volume} {44}},\ \bibinfo {pages} {131} (\bibinfo {year} {2001})}\BibitemShut {NoStop}%
\bibitem [{\citenamefont {S{\'e}n{\'e}chal}(2004)}]{Senechal2004}%
  \BibitemOpen
  \bibfield  {author} {\bibinfo {author} {\bibfnamefont {D.}~\bibnamefont {S{\'e}n{\'e}chal}},\ }\bibinfo {title} {An introduction to bosonization},\ in\ \href {https://doi.org/10.1007/0-387-21717-7{\_}4} {\emph {\bibinfo {booktitle} {Theoretical Methods for Strongly Correlated Electrons}}},\ \bibinfo {editor} {edited by\ \bibinfo {editor} {\bibfnamefont {D.}~\bibnamefont {S{\'e}n{\'e}chal}}, \bibinfo {editor} {\bibfnamefont {A.-M.}\ \bibnamefont {Tremblay}},\ and\ \bibinfo {editor} {\bibfnamefont {C.}~\bibnamefont {Bourbonnais}}}\ (\bibinfo  {publisher} {Springer New York},\ \bibinfo {address} {New York, NY},\ \bibinfo {year} {2004})\ pp.\ \bibinfo {pages} {139--186}\BibitemShut {NoStop}%
\bibitem [{\citenamefont {Japaridze}\ and\ \citenamefont {M\"uller-Hartmann}(2000)}]{Japaridze2000tsi}%
  \BibitemOpen
  \bibfield  {author} {\bibinfo {author} {\bibfnamefont {G.~I.}\ \bibnamefont {Japaridze}}\ and\ \bibinfo {author} {\bibfnamefont {E.}~\bibnamefont {M\"uller-Hartmann}},\ }\bibfield  {title} {\bibinfo {title} {Triplet superconductivity in a one-dimensional ferromagnetic $t\ensuremath{-}{J}$ model},\ }\href {https://doi.org/10.1103/PhysRevB.61.9019} {\bibfield  {journal} {\bibinfo  {journal} {Phys. Rev. B}\ }\textbf {\bibinfo {volume} {61}},\ \bibinfo {pages} {9019} (\bibinfo {year} {2000})}\BibitemShut {NoStop}%
\bibitem [{\citenamefont {Dziurzik}\ \emph {et~al.}(2004)\citenamefont {Dziurzik}, \citenamefont {Japaridze}, \citenamefont {Schadschneider},\ and\ \citenamefont {Zittartz}}]{Dziurzik2004tsv}%
  \BibitemOpen
  \bibfield  {author} {\bibinfo {author} {\bibfnamefont {C.}~\bibnamefont {Dziurzik}}, \bibinfo {author} {\bibfnamefont {G.~I.}\ \bibnamefont {Japaridze}}, \bibinfo {author} {\bibfnamefont {A.}~\bibnamefont {Schadschneider}},\ and\ \bibinfo {author} {\bibfnamefont {J.}~\bibnamefont {Zittartz}},\ }\bibfield  {title} {\bibinfo {title} {Triplet superconductivity vs. easy-plane ferromagnetism in a {1D} itinerant electron system with transverse spin anisotropy},\ }\href {https://doi.org/10.1140/epjb/e2004-00081-5} {\bibfield  {journal} {\bibinfo  {journal} {The European Physical Journal B - Condensed Matter and Complex Systems}\ }\textbf {\bibinfo {volume} {37}},\ \bibinfo {pages} {453} (\bibinfo {year} {2004})}\BibitemShut {NoStop}%
\bibitem [{\citenamefont {Hauschild}\ and\ \citenamefont {Pollmann}(2018)}]{Hauschild2018Ens}%
  \BibitemOpen
  \bibfield  {author} {\bibinfo {author} {\bibfnamefont {J.}~\bibnamefont {Hauschild}}\ and\ \bibinfo {author} {\bibfnamefont {F.}~\bibnamefont {Pollmann}},\ }\bibfield  {title} {\bibinfo {title} {{Efficient numerical simulations with Tensor Networks: Tensor Network Python (TeNPy)}},\ }\href {https://doi.org/10.21468/SciPostPhysLectNotes.5} {\bibfield  {journal} {\bibinfo  {journal} {SciPost Phys. Lect. Notes}\ ,\ \bibinfo {pages} {5}} (\bibinfo {year} {2018})},\ \bibinfo {note} {code available from \url{https://github.com/tenpy/tenpy}},\ \Eprint {https://arxiv.org/abs/1805.00055} {arXiv:1805.00055} \BibitemShut {NoStop}%
\end{thebibliography}%
\end{document}